\documentclass[aps,prd,twocolumn,floatfix,superscriptaddress,preprintnumber,amsmath,amssymb,eqsecnum,showkeys,nofootinbib]{revtex4-1}
\pdfoutput=1
\usepackage{subfigure}
\usepackage{graphicx}
\usepackage{xcolor}
\newcommand{\be}{\begin{equation}}
\newcommand{\ee}{\end{equation}}
\newcommand{\bea}{\begin{eqnarray}}
\newcommand{\eea}{\end{eqnarray}}

\begin{document}
\title{Asymmetric Dark Matter in the Sun: A Multicomponent Perspective}

\author{Amit Dutta Banik}
\email[E-mail: ]{amitdbanik@gmail.com}
\affiliation{Physics and Applied Mathematics Unit, Indian Statistical Institute, Kolkata-700108, India}

\date{\today}

\begin{abstract}
We present a novel concept of enhanced asymmetric dark matter annihilation in astrophysical bodies like the Sun in the presence of multiple dark matter candidates based on hidden annihilation mechanisms. We consider hidden sector annihilation of a heavy dark matter into an asymmetric dark matter, resulting in a significant change in the dark matter annihilation flux and the muon flux at neutrino detectors. We quantify expected changes in the muon flux with scaling parameters for the symmetric or asymmetric nature of the heavier dark matter candidate.
  
\end{abstract}

\maketitle

\section {Introduction}
\label{Intro}

\noindent The overwhelming presence of dark matter (DM) in the universe is ensured by gravitational and cosmological observations~\cite{Aghanim:2018eyx}. Although being five times more abundant than the ordinary matter constituent of the universe, its very nature in the grand cosmos is unknown to human perception, as it still evades detection with conventional experiments performed via direct and indirect detection. To perceive the nature of particle-like dark matter, the concept of weakly interacting massive particles (WIMPs) is proposed and widely accepted in particle physics.
WIMPs are considered to be beyond Standard Model (BSM) particles that are cold relics due to the thermal freeze-out of WIMP interactions in the expanding universe~\cite{Jungman:1995df,Bertone:2004pz}. Direct detection of WIMPs can be probed through their interaction and scattering with ordinary matter, performed by various direct search experiments. The indirect detection probe of WIMPs searches for the dark matter annihilation signatures from observations at possible sources like the galactic centre, dwarf galaxies, and compact astrophysical objects like stars, white dwarfs, and neutron stars that can easily capture DM particles with their strong gravity. The direct detection experiments XENON1T \cite{Aprile:2018dbl,Aprile:2015uzo}, XENONnT \cite{XENON:2020kmp,XENON:2023sxq}, PandaX-II \cite{Cui:2017nnn}, PICO \cite{PICO:2019vsc} provide constraints on the scattering cross-sections of dark matter, and indirect detection experiments like Fermi-LAT \cite{Fermi-LAT:2015att}, DES \cite{Fermi-LAT:2015ycq}, H.E.S.S. \cite{HESS:2016mib}, MAGIC \cite{MAGIC:2016xys} constrain the DM annihilation cross-sections based on the observed flux of produced SM particles.
Indirect detections, similar to direct detection, can also constrain DM scattering cross-sections from the study of capture and annihilation of dark matter in compact objects.
Based on the observed muon neutrino flux that can be produced from DM annihilation in the Sun, neutrino detector experiments like IceCube \cite{IceCube:2016dgk}, Super-K \cite{Super-Kamiokande:2015xms}, ANTARES \cite{ANTARES:2016xuh}, etc., limit the DM scattering cross-sections. 

\noindent Experimental constraints on DM scattering and annihilation properties derived from direct and indirect detection experiments are based on the simple assumption that there exists only one type of dark matter candidate. However, similar to the Standard Model, the dark sector can also be comprised of multiple candidates.
Apart from that, the existence of an asymmetry in visible matter with an overabundance of particles over antiparticles 
also compels us to investigate whether the dark sector also possesses a similar asymmetry. The concept of multiple dark matter comes with a rich phenomenology and also challenges the present methods of direct and indirect detection of dark matter. In the case of multicomponent dark matter with WIMP-like particles, thermal evolution of dark matter candidates in the early universe gets affected due to the annihilation between the candidates. A similar situation occurs with the late-time study of dark matter candidates as they get captured and annihilated in compact astrophysical bodies like the Sun. The indirect detection signature of the DM candidate inside the Sun for the single DM component is explored in many literature~\cite {Faulkner:1985rm,Griest:1986yu,Gould:1987ju,Gould:1991hx,Barger:2007xf,Hooper:2008cf,Belotsky:2008vh,Wikstrom:2009kw,Erkoca:2009by,Zentner:2009is,Covi:2009xn,Chen:2011vda,Bernal:2012qh,Chen:2014oaa,Catena:2016ckl,Tiwari:2018gxz,Gaidau:2018yws,Gupta:2022lws}. Although an ample amount phenomenological studies are performed with multi-particle dark matter~\cite{Belanger:2011ww,Khlopov:1995pa,Bhattacharya:2013hva,Bian:2013wna,Esch:2014jpa,Ahmed:2017dbb,Herrero-Garcia:2017vrl,Herrero-Garcia:2018qnz, Aoki:2018gjf, Bhattacharya:2016ysw,Chakraborti:2018aae,Elahi:2019jeo,Borah:2019aeq,Bhattacharya:2019tqq,DuttaBanik:2020jrj,DiazSaez:2021pfw,BasiBeneito:2022qxd,Elor:2020tkc,Elahi:2021jia}, only a few of the works visualised the concept of indirect signature of multiple dark matter evolution inside the Sun ~\cite{Aoki:2012ub,Aoki:2014lha,Berger:2014sqa,Aoki:2017eqn,DuttaBanik:2023yxj}. 
In the present work, we investigate the evolution of asymmetric dark matter (ADM) inside the Sun, accompanied by a heavier DM candidate. The annihilation of heavier candidates into ADM modifies the dynamics of capture and annihilation of the said ADM. The changes in the evolution of the ADM will depend on the symmetric or asymmetric nature of other dark matter. In this work, we report possible deviations in asymmetric dark matter annihilation in the Sun in the presence of internal conversion between dark matter candidates and its impact on the detection of ADM at neutrino detectors. 

\noindent The paper is organized as follows. In Sec.~\ref{BE}, we construct the Boltzmann equations for multi-particle dark matter with asymmetric dark matter. The dynamics of dark matter capture and annihilation in the Sun is established in Sec.~\ref{Sun}, and the detection prospect of asymmetric dark matter is presented in Sec.~\ref{res}. In Sec.~\ref{con}, we summarize the work with concluding remarks. 

\section{Boltzmann equations for coupled asymmetric dark matter}
\label{BE}
\noindent Profound studies on the single component asymmetric dark matter are conducted in many literatures \cite{Kaplan:2009ag,Graesser:2011wi,Iminniyaz:2011yp,Gelmini:2013awa,Zurek:2013wia,Kitabayashi:2015oda,Blennow:2015xha,Iminniyaz:2016iom,Agrawal:2016uwf,Nagata:2016knk,Baldes:2017gzw,Gresham:2017cvl,HajiSadeghi:2017zrl,Murase:2016nwx,Gresham:2018anj,Lopes:2019jca,Ghosh:2020lma,Ghosh:2021qbo,Ho:2022erb,Steigerwald:2022pjo,Ho:2022tbw,Roy:2024ear}. In this section, we venture into the coupled dark matter scenarios where the lighter DM candidate is recognised as an asymmetric dark matter.
The aforementioned ADM above leads to two possible coupled DM scenarios: I) one symmetric dark matter coupled to another asymmetric dark matter, and II) two different asymmetric dark matter candidates coupled to each other. We consider both DM candidates to be stable and protected by some symmetries ($Z_2,~Z_4$ etc.). With the consideration above, we derive Boltzmann equations for dark matter evolution in the early universe in a model agnostic approach. The coupled symmetric and asymmetric DM scenario is proposed in recent work \cite{DuttaBanik:2024vro}. The corresponding coupled Boltzmann equations for a heavy symmetric dark matter $S$ with a companion asymmetric dark matter $A$ ($m_S>m_{A}=m_{\bar A}$) are expressed in terms of their co-moving densities ($Y_k=n_k/s;k=S,A,\bar A$) as 
{\small
\begin{align} \label{eq:boltzmann1}
  \frac{Hx}{s}\frac{dY_A}{dx}&= -{\langle \sigma v \rangle}_{A\bar{A}} \left(Y_A Y_{\bar{A}}-Y_{A,\rm eq}Y_{\bar{A},\rm eq}\right) \nonumber \\ &+ {\langle \sigma v \rangle}_{SS \rightarrow A\bar{A}} \left(Y_S^2-\frac{Y_A Y_{\bar{A}}}{Y_{A,\rm eq}Y_{\bar{A},\rm eq}}Y_{S,\rm eq}^{2}\right)\, ,
\nonumber\\
  \frac{Hx}{s}\frac{dY_{\bar{A}}}{dx}&= -{\langle \sigma v \rangle}_{A \bar{A}} \left(Y_A Y_{\bar{A}}-Y_{A,\rm eq}Y_{\bar{A},\rm eq}\right) \nonumber \\  &+ {\langle \sigma v \rangle}_{SS \rightarrow A\bar{A}} \left(Y_S^2-\frac{Y_A Y_{\bar{A}}}{Y_{A,\rm eq}Y_{\bar{A},\rm eq}}Y_{S,\rm eq}^{2}\right)\, ,
\nonumber\\
 \frac{Hx}{s}\frac{dY_S}{dx} &= -{\langle \sigma v \rangle}_{S S} \left(Y_S^2-Y_{S,\rm eq}^2\right) \nonumber \\ &
-2{\langle \sigma v \rangle}_{SS \rightarrow A\bar{A}} \left(Y_S^2-\frac{Y_A Y_{\bar{A}}}{Y_{A,\rm eq}Y_{\bar{A},\rm eq}}Y_{S,\rm eq}^{2}\right), 
\end{align}
}
\noindent where $x=m_S/T$, and $H$ is the Hubble expansion rate. In Eq.~(\ref{eq:boltzmann1}) above,
${\langle \sigma v \rangle}_{A\bar A}$ and ${\langle \sigma v \rangle}_{S S}$ denote the annihilation of DM candidates into SM particles, whereas the exchange annihilation between dark matter particles is given by ${\langle \sigma v \rangle}_{SS \rightarrow A\bar{A}}$. The above set of coupled Boltzmann equations are valid in the absence of self-annihilation of asymmetric DM $AA\rightarrow {\rm SM SM},~SS$ and  $\bar A \bar A\rightarrow {\rm SMSM},~SS$.
Comoving number densities of dark sector particles in equilibrium are given as
\begin{align}\label{yeq}
Y_{A,\rm eq}=\frac{g_A}{s}\left(\frac{m_{A}T}{2\pi}\right)^{3/2}e^{(-m_A+\mu_{A})/T},\nonumber \\
Y_{\bar{A},\rm eq}=\frac{g_A}{s}\left(\frac{m_{A}T}{2\pi}\right)^{3/2}e^{(-m_A-\mu_{A})/T},\nonumber \\
Y_{S,\rm eq}=\frac{g_S}{s}\left(\frac{m_{S}T}{2\pi}\right)^{3/2}e^{-m_S/T}\, ,
\end{align}
with $s$ being the entropy density, and the internal degrees of freedom of a species is
denoted by $g_i,~i=S,A$ with $g_{A}=g_{\bar A}$. Throughout the evolution, the asymmetric DM candidate respects the following condition
\be
\frac{d(Y_{A}-Y_{\bar{A}})}{dx}=0\, ,
\ee
indicating that $Y_A-Y_{A}=C$, is constant. Therefore, if a net asymmetry is created in the early epoch between $Y_A$ and $Y_{A}$, it survives when the self-annihilation of asymmetric DM is absent. This asymmetry can be generated at a very high energy scale through the decay of beyond SM particles or via inflation that does not affect the freeze-out process of DM happening at a much lower energy scale. Therefore, Eq.~(\ref{eq:boltzmann1})
can be rearranged as 
{\small
\begin{align} \label{eq:boltzmann2}
  \frac{Hx}{s}\frac{dY_A}{dx}&= -{\langle \sigma v \rangle}_{A \bar{A}} \left(Y_A^2-CY_A-Y_{A,\rm eq}Y_{\bar{A},\rm eq}\right) \nonumber \\&+ {\langle \sigma v \rangle}_{SS \rightarrow A\bar{A}} \left(Y_S^2-\frac{Y_A^2-CY_A}{Y_{A,\rm eq}Y_{\bar{A},\rm eq}}Y_{S,\rm eq}^{2}\right)\, ,
\nonumber\\
  \frac{Hx}{s}\frac{dY_{\bar{A}}}{dx}&= -{\langle \sigma v \rangle}_{A \bar{A}} \left(Y_{\bar{A}}^2+CY_{\bar{A}}-Y_{A,\rm eq}Y_{\bar{A},\rm eq}\right)\nonumber \\& + {\langle \sigma v \rangle}_{SS \rightarrow A\bar{A}} \left(Y_S^2-\frac{Y_{\bar{A}}^2+CY_{\bar{A}}}{Y_{A,\rm eq}Y_{\bar{A},\rm eq}}Y_{S,\rm eq}^{2}\right)\, ,
\nonumber\\
 \frac{Hx}{s}\frac{dY_S}{dx} &= -{\langle \sigma v \rangle}_{S S} \left(Y_S^2-Y_{S,\rm eq}^2\right) \nonumber \\&
-{\langle \sigma v \rangle}_{SS \rightarrow A\bar{A}} \left(Y_S^2-\frac{Y_{\bar{A}}^2+CY_{\bar{A}}}{Y_{A,\rm eq}Y_{\bar{A},\rm eq}}Y_{S,\rm eq}^{2}\right)
\nonumber\\
& \qquad -{\langle \sigma v \rangle}_{SS \rightarrow A\bar{A}} \left(Y_S^2-\frac{Y_A^2-CY_A}{Y_{A,\rm eq}Y_{\bar{A},\rm eq}}Y_{S,\rm eq}^{2}\right). 
\end{align}
}
\noindent The asymmetry parameter $C$ can be defined in terms of the chemical potential $\mu_A$ \cite{Gelmini:2013awa}
\begin{equation}
e^{\mu_A /T}=\frac{1}{2}\left(\frac{Cs}{n_{A,\rm eq}(\mu =0)}+\sqrt{4+\left(\frac{Cs}{n_{A,\rm eq}(\mu =0)}\right)^2}\right)\, ,
\label{muandC}
\end{equation}
and relic abundances of different dark matter candidates are obtained by solving for the coupled Boltzmann equations
\begin{gather}
  \Omega_k {\rm h}^2=2.755\times10^8\frac{m_k}{GeV} \, Y_k(T_0) \, ,
   \quad k=S,(A,\bar{A})\, .
  \label{relic}
\end{gather}
The DM components must account for the total DM relic abundance obtained by the Planck experiment \cite{Aghanim:2018eyx}
\begin{gather}
  \Omega_S {\rm h}^2+\Omega_A {\rm h}^2+\Omega_{\bar A} {\rm h}^2=\Omega_{\rm DM}{\rm h}^2=0.1199\pm0.0019 \, .
  \label{relic2}
\end{gather}
Let us consider that the symmetric DM candidate shares a fraction $f$ of the total DM relic density. Therefore, the components of asymmetric DM share $(1-f)$ fraction of the total DM abundance. The fractional asymmetry between the components of $A$ and $\bar A$ is defined as $r=\frac{Y_{\bar A}(T_0)}{Y_A(T_0)}=\frac{n_{\bar A}(T_0)}{n_A(T_0)}$, with $T_0$ being the present day temperature. 
Hence, for $C>0$, one finds
\begin{gather}
  \Omega_A {\rm h}^2=\frac{(1-f)}{(1+r)}\Omega_{\rm DM}{\rm h}^2\, , \hskip 5mm \Omega_{\bar A} {\rm h}^2=\frac{(1-f)r}{(1+r)}\Omega_{\rm DM}{\rm h}^2\, .
  \label{relic2a}
\end{gather}
In a standard single-component ADM scenario, the fractional asymmetry parameter $r$ is related to $C$, DM annihilation cross-section and the relic abundances of its components~\cite{Gelmini:2013awa}. However, the coupled nature of BEs removes their correlations and sets them as free parameters. This allows a broader range of phenomenological prospects, relaxing the parameters involved in conventional treatments of ADM.
\noindent 
Apart from the multicomponent DM scenario with admixture of symmetric and asymmetric DM components, we also consider the case of coupled multiple asymmetric dark matter candidates. We consider two dark matter candidates with particles $A$ and $B$ and anti-particles $\bar A$ and $\bar B$. Both dark matter candidates having a finite asymmetry, new coupled Boltzmann equations are described as
{\small
\begin{align} \label{eq:boltzmann5}
  \frac{Hx}{s}\frac{dY_A}{dx}= -{\langle \sigma v \rangle}_{A\bar{A}} \left(Y_A^2-C_AY_A-Y_{A,\rm eq}Y_{\bar{A},\rm eq}\right) \nonumber \\- {\langle \sigma v \rangle}_{A\bar A \rightarrow B\bar{B}} \left(Y_A^2-C_AY_A-\frac{Y_B^2-C_BY_B}{Y_{B,\rm eq}Y_{\bar{B},\rm eq}}Y_{A,\rm eq}Y_{\bar{A},\rm eq}\right)\,,
\nonumber\\
  \frac{Hx}{s}\frac{dY_{\bar{A}}}{dx}= -{\langle \sigma v \rangle}_{A\bar{A}} \left(Y_{\bar{A}}^2+C_AY_{\bar{A}}-Y_{A,\rm eq}Y_{\bar{A},\rm eq}\right) \nonumber \\- {\langle \sigma v \rangle}_{A\bar A \rightarrow B\bar{B}} \left(Y_{\bar{A}}^2+C_AY_{\bar{A}}-\frac{Y_{\bar{B}}^2+C_BY_{\bar{B}}}{Y_{B,\rm eq}Y_{\bar{B},\rm eq}}Y_{A,\rm eq}Y_{\bar{A},\rm eq}\right)\,,
\nonumber\\
\frac{Hx}{s}\frac{dY_B}{dx}= -{\langle \sigma v \rangle}_{B\bar{B}} \left(Y_B^2-C_BY_B-Y_{B,\rm eq}Y_{\bar{B},\rm eq}\right) \nonumber \\+ {\langle \sigma v \rangle}_{A\bar A \rightarrow B\bar{B}} \left(Y_A^2-C_AY_A-\frac{Y_B^2-C_BY_B}{Y_{B,\rm eq}Y_{\bar{B},\rm eq}}Y_{A,\rm eq}Y_{\bar{A},\rm eq}\right)\,,
\nonumber\\
  \frac{Hx}{s}\frac{dY_{\bar{B}}}{dx}= -{\langle \sigma v \rangle}_{B\bar{B}} \left(Y_{\bar{B}}^2+C_BY_{\bar{B}}-Y_{B,\rm eq}Y_{\bar{B},\rm eq}\right) \nonumber \\+ {\langle \sigma v \rangle}_{A\bar A \rightarrow B\bar{B}} \left(Y_{\bar{A}}^2+C_AY_{\bar{A}}-\frac{Y_{\bar{B}}^2+C_BY_{\bar{B}}}{Y_{B,\rm eq}Y_{\bar{B},\rm eq}}Y_{A,\rm eq}Y_{\bar{A},\rm eq}\right)\,. 
\end{align}
}

\noindent 
In the above Eq.~(\ref{eq:boltzmann5}), we impose two asymmetry parameters $C_A=Y_A-Y_{\bar A}$ and $C_B=Y_B-Y_{\bar B}$ such that $\frac{d(Y_{k}-Y_{\bar{k}})}{dx}=0; k=A,B$ is satisfied. The asymmetry parameters $C_A,~C_B$ can be related to the chemical potential $\mu_A,~\mu_B$ of DM candidates in the form of Eq.~(\ref{muandC}). The DM abundance with two asymmetric DM sectors reads as

\begin{gather}
  \Omega_A {\rm h}^2+\Omega_{\bar A} {\rm h}^2+\Omega_B {\rm h}^2+\Omega_{\bar B} {\rm h}^2=\Omega_{\rm DM}{\rm h}^2\, .
  \label{relic3}
\end{gather}

\noindent Eq.~(\ref{relic3}) can be simplified as
\begin{gather}
  \Omega_A {\rm h}^2(1+r_A)+\Omega_B {\rm h}^2(1+r_B)=\Omega_{\rm DM}{\rm h}^2\,,
  \label{relic4}
\end{gather}
\noindent where $r_A$ and $r_B$ represent the fractional asymmetry of species $A$ and $B$ at present. For $\Omega_A {\rm h}^2(1+r_A)=f\Omega_{\rm DM}{\rm h}^2$ one can express the relic densities of different components as
\begin{gather}
  \Omega_A {\rm h}^2=\frac{f}{1+r_A}\Omega_{\rm DM}{\rm h}^2\, ,\hskip 5mm 
  \Omega_{\bar A} {\rm h}^2=\frac{r_Af}{1+r_A}\Omega_{\rm DM}{\rm h}^2\, \nonumber \\
  \Omega_B {\rm h}^2=\frac{1-f}{1+r_B}\Omega_{\rm DM}{\rm h}^2\, ,\hskip 5mm
  \Omega_{\bar B} {\rm h}^2=\frac{(1-f)r_B}{1+r_B}\Omega_{\rm DM}{\rm h}^2\, .\hskip 5mm 
  \label{relic5}
\end{gather}

\section{ADM Capture and annihilation in the Sun}
\label{Sun}
\noindent In this section, we encapsulate the dark matter capture and annihilation dynamics inside the Sun.
Dark matter candidates (symmetric and asymmetric components) are captured by the Sun and undergo annihilation into the SM sector and the hidden sector. For $m_S>m_{A,\bar A}$, conversion $SS\rightarrow A \bar A$ can take place, leading to significant deviation from the conventional dynamics of ADM \cite{Murase:2016nwx}. The evolution rate of the DM components inside the Sun is given as

\begin{align}
&\frac{dN_S}{dt} = fC^c_{S}  -C^{ann}_{SS} N_S^{2}-C^{ann}_{SSA\bar A}  N_S^{2}\, , \nonumber \\
&\frac{dN_A}{dt} = \frac{(1-f)}{1+r}C^c_{A}  -C^{ann}_{A\bar A}  N_A N_{\bar A}+\gamma C^{ann}_{SSA\bar A}  N_S^{2}\, , \nonumber \\
&\frac{dN_{\bar A}}{dt} = \frac{(1-f)r}{1+r}C^c_{\bar A}  -C^{ann}_{A\bar A}  N_A N_{\bar A}+\gamma C^{ann}_{SSA\bar A}  N_S^{2}\, ,
\label{3}
\end{align} 
where $C^c_k; k=S,A,\bar A$ is the capture rate coefficient of a DM species and $C^{ann}_{kk,k\bar k}$ is the annihilation rate coefficient of the DM species into the SM sector. Since components $A$ and $\bar A$ have identical scattering interactions, $C^c_{\bar A}=C^c_A$ is assumed. The exchange annihilation rate between the DM candidates is expressed by the coefficient $C^{ann}_{SSAA}$. Annihilation coefficients in Eq.~(\ref{3}) are given as
\begin{gather}
C^{ann}_{SS}= \left\langle \sigma v \right\rangle_{SS}\frac{V_{2S}}{V_{1S}^{2}}\, , \hskip 5mm 
C^{ann}_{A\bar A}= \left\langle \sigma v \right\rangle_{A\bar A}\frac{V_{2A}}{V_{1A}^{2}}\, , \nonumber \\  
C^{ann}_{SSA\bar A} = \left\langle \sigma v \right\rangle_{SS\rightarrow A\bar A}\frac{V_{2S}}{V_{1S}^{2}}\, ,
\label{C12}
\end{gather}
where the volume term is given as
\begin{equation}
V_{jk}\simeq6.5\times10^{28}\textrm{ cm}^{3}\left(\frac{10\textrm{ GeV}}{jm_{k}}\right)^{3/2}\, ,~k=S,A.
\label{vol}
\end{equation}

\noindent The ADM production term in Eq.~(\ref{3}) through the annihilation process $SS\rightarrow A\bar A$ is also accompanied by a coefficient of absorption $\gamma$ that signifies the amount of produced $A,~\bar A$ particles that fail to escape the Sun due to interaction.  We consider $\gamma\simeq 0.05$ to be small but non-negligible due to small mass splitting of DM candidates and a fraction of DM produced gets trapped as they thermalise after depositing their energy in the solar medium upon interaction \footnote{Produced dark mater particles can also undergo deep inelastic scattering (DIS) and transfer a significant amount of energy into the stellar medium~\cite{Su:2024flx,Wang:2025ztb}. Therefore, a small but significant part of DM particles produced from dark sector annihilation may end up cpatured in the Sun.}. 
The effect of evaporation of produced the DM is taken into consideration in a recent work \cite{DuttaBanik:2023yxj} that was overlooked in the previous study \cite{Aoki:2012ub}.
Note that $\gamma$ can be different for different celestial bodies, for example, $\gamma$ can be close to unity for neutron stars having strong gravitational pull and dense medium.

\noindent Turning into the case where both the dark matter candidates are asymmetric that follow the solution to Boltzmann Eq.~(\ref{eq:boltzmann5}), the evolution of DM particles inside the Sun is defined as
\begin{align}
&\frac{dN_A}{dt} = \frac{f}{1+r_A}C^c_{A}  -C^{ann}_{A\bar A}  N_A N_{\bar A}- C^{ann}_{A\bar AB\bar B}  N_AN_{\bar A}\, , \nonumber \\
&\frac{dN_{\bar A}}{dt} = \frac{fr_A}{1+r_A}C^c_{\bar A}  -C^{ann}_{A\bar A}  N_AN_{\bar A}-C^{ann}_{A\bar AB\bar B}  N_AN_{\bar A}\, ,\nonumber \\
&\frac{dN_B}{dt} = \frac{(1-f)}{1+r_B}C^c_{B}  -C^{ann}_{B\bar B}  N_B N_{\bar B}+\gamma C^{ann}_{A\bar AB\bar B}   N_AN_{\bar A}\, , \nonumber \\
&\frac{dN_{\bar B}}{dt} = \frac{(1-f)r_B}{1+r_B}C^c_{\bar B}  -C^{ann}_{B\bar B}  N_BN_{\bar B}+\gamma C^{ann}_{A\bar AB \bar B}   N_AN_{\bar A}\, ,
\label{4}
\end{align} 
with the annihilation rate coefficients for DM candidates given as 
\begin{gather}
C^{ann}_{A{\bar A}}= \left\langle \sigma v \right\rangle_{A\bar A}\frac{V_{2A}}{V_{1A}^{2}}\, , \hskip 5mm 
C^{ann}_{B\bar B}= \left\langle \sigma v \right\rangle_{B\bar B}\frac{V_{2B}}{V_{1B}^{2}}\, , \nonumber \\  
C^{ann}_{A\bar AB\bar B} = \left\langle \sigma v \right\rangle_{A\bar A\rightarrow B\bar B}\frac{V_{2A}}{V_{1A}^{2}}\, ,
\label{C12a}
\end{gather}
and capture rates follow the relation $C^c_{k}=C^c_{\bar k};~k=A,B$. In the present work, we consider dark matter spin-dependent interaction only. Spin-dependent interaction of DM can be obtained via axial vector type interactions of dark matter candidates. The interaction term of the form  $\bar f_{\rm DM}\gamma_\mu\gamma_5f_{\rm DM}\bar f_{\rm SM}\gamma_\mu\gamma_5f_{\rm SM}$ is responsible for spin-dependent dark matter scattering. Mixing between DM components through the interaction term $\bar f^{\prime}_{\rm DM}\gamma_\mu\gamma_5f_{\rm DM}\bar f_{\rm SM}\gamma_\mu\gamma_5f_{\rm SM}$ ($f_{\rm DM}$ and $f^{\prime}_{\rm DM}$ being different DM species) is prohibited by the symmetry arguments of DM candidates. With the above specifications, the DM annihilation cross sections appearing in various annihilation rate coefficients mentioned in Eqs.~(\ref{C12},~\ref{C12a}) into SM and hidden sector are not velocity or momentum suppressed. The dark matter capture rate coefficient for spin-dependent interaction of DM can be expressed as ~\cite{Jungman:1995df,Bertone:2004pz}
\begin{align}
C_{\text{c}}\simeq3.35\times10^{24}\textrm{ s}^{-1}\left(\frac{\rho_{0}}{0.3\textrm{ GeV/cm}^{3}}\right)\left(\frac{270\textrm{ km/s}}{\bar{v}}\right)^{3} \nonumber \\
\left(\frac{\textrm{GeV}}{m_{\chi}}\right)^{2}\left(\frac{\sigma_{\textrm{H}}^{\textrm{SD}}}{10^{-6}\textrm{ pb}}\right).
\label{eq:capture_SD}
\end{align}
with $\sigma_{\rm H}^{\textrm{SD}}$ of the form 
\begin{equation}
\sigma_{\rm H}^{\textrm{SD}}=
\frac{4(J+1)}{3 J}\left|\left\langle S_{p}\right\rangle +\left\langle S_{n}\right\rangle \right|^{2}
\sigma_{\chi p}^{\textrm{SD}}\, .
\end{equation}
Using the expressions above, the population of dark matter candidates $N_S,~N_{A,\bar A},~N_{B,\bar B}$ inside the Sun, described in Eq.~(\ref{3},~\ref{4}), can be obtained with known values of DM masses, scattering cross-sections, and their annihilation cross-sections into SM and dark sector accompanied by absorption coefficient, fractional abundance and fractional asymmetry parameters.

\section{Results}
\label{res}

\begin{figure}
    \begin{center}
        \includegraphics[width=0.45\textwidth]{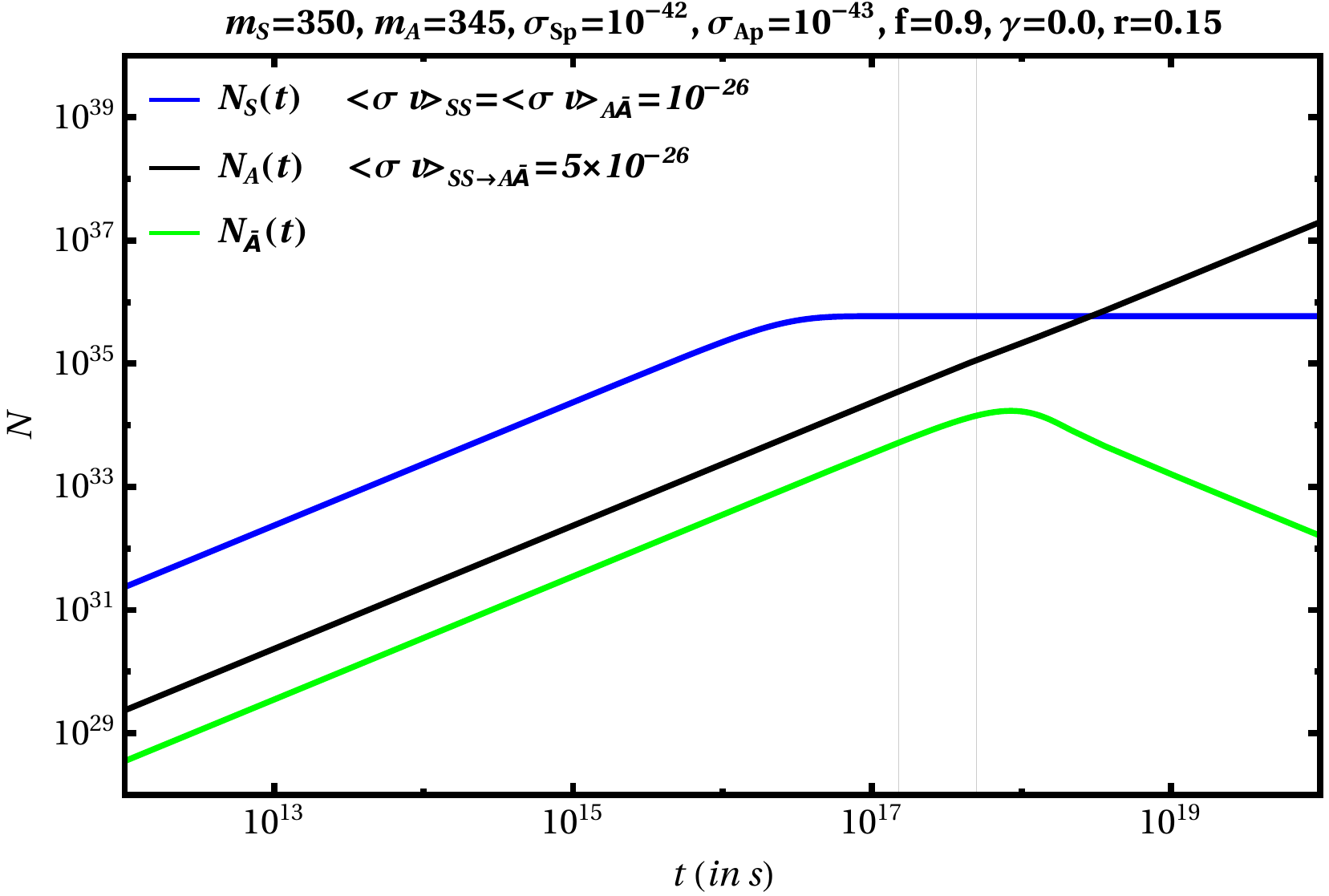}
        \includegraphics[width=0.45\textwidth]{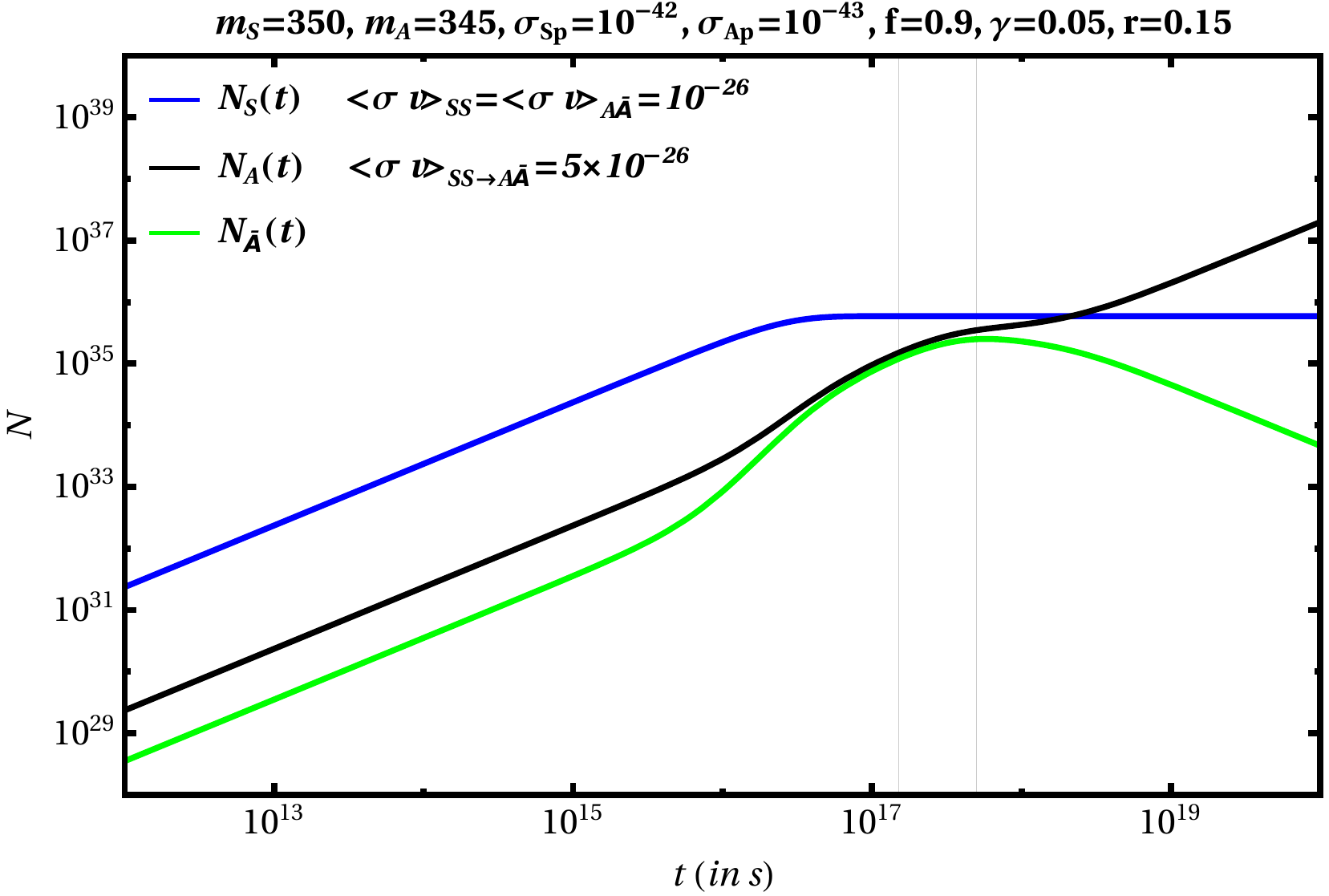}
        \includegraphics[width=0.45\textwidth]{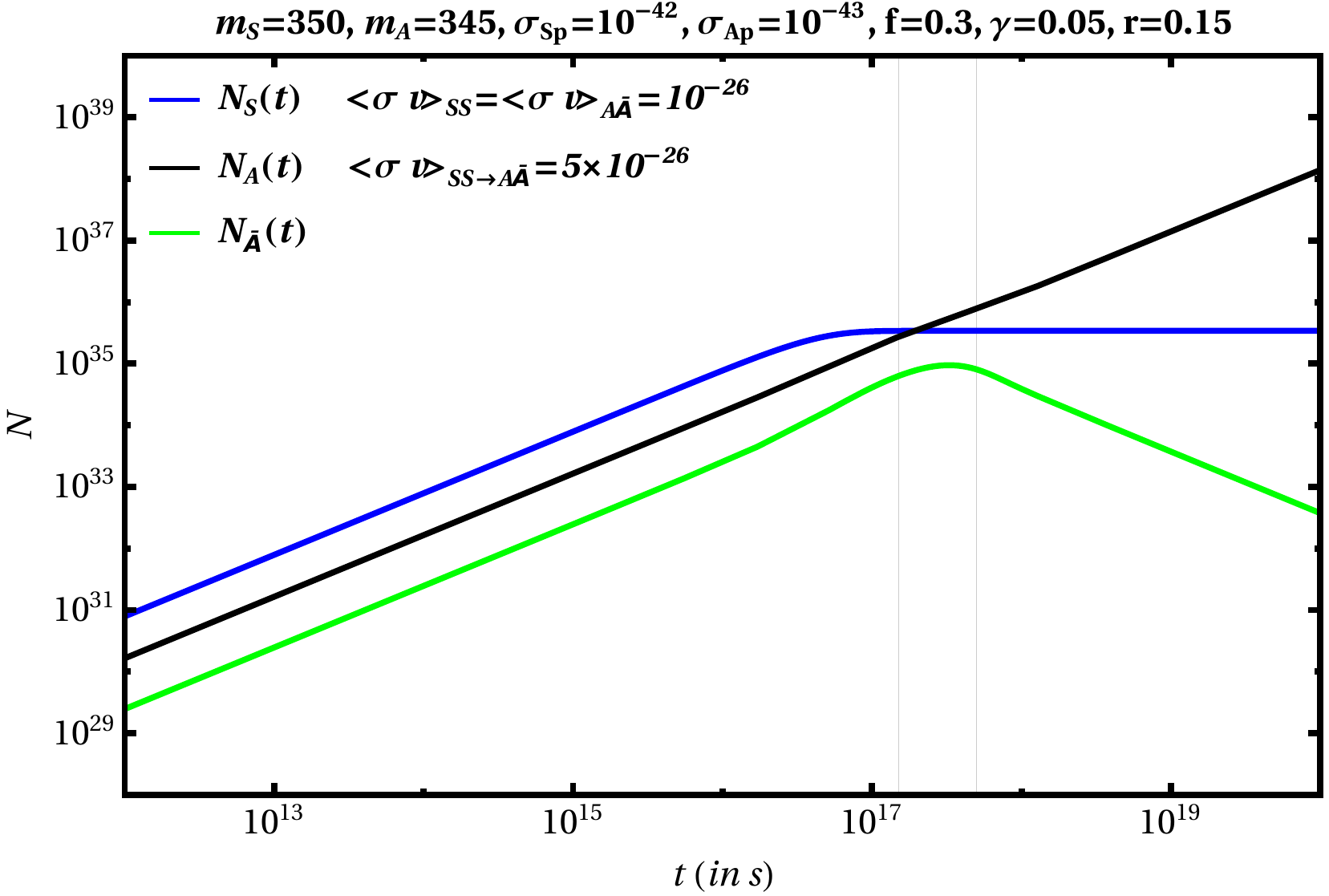}
        \includegraphics[width=0.45\textwidth]{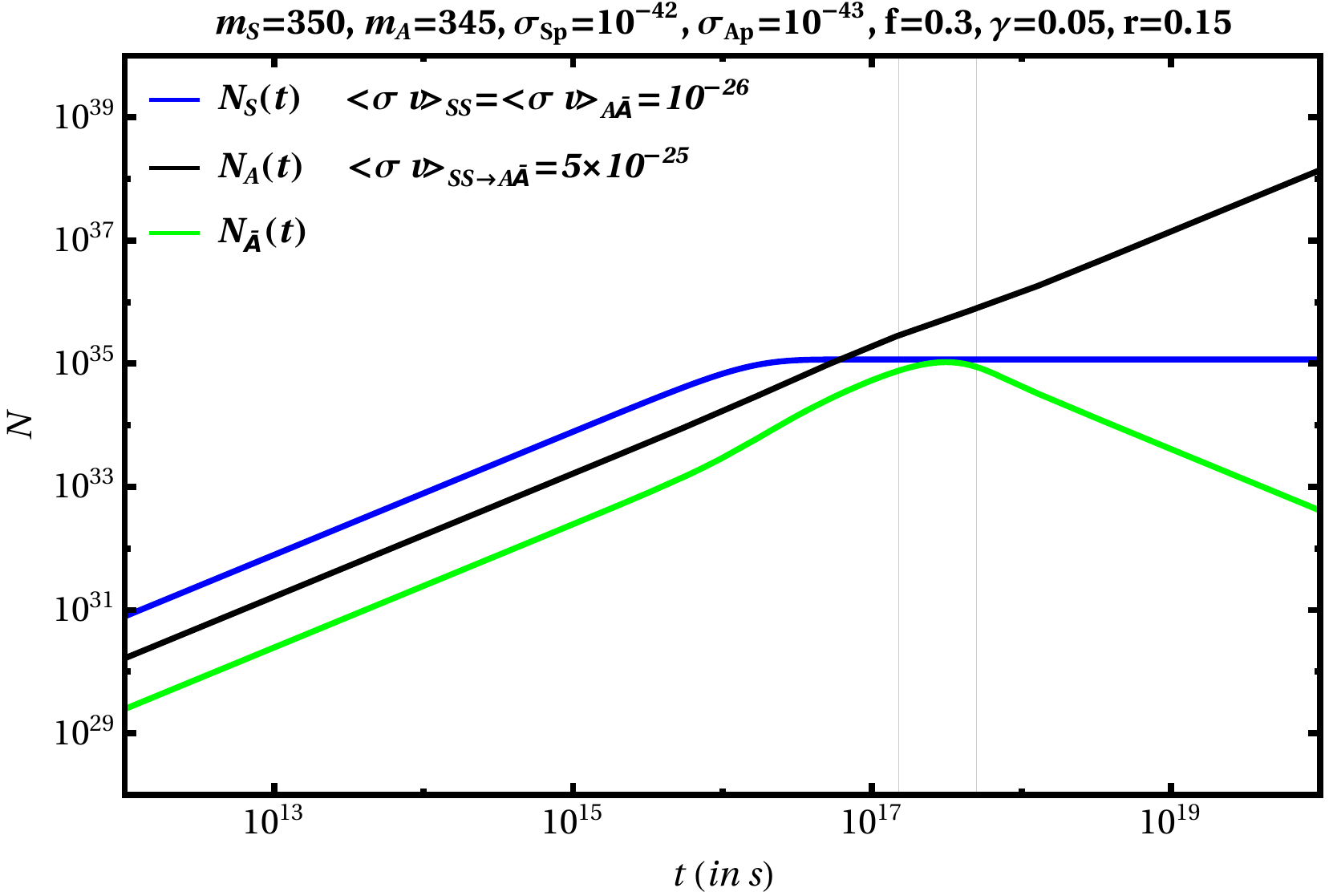}
            \caption{\it Time evolution of $N_S$ and $N_{A,\bar A}$ population inside the Sun for variation of $\gamma$, $f$, and $\left\langle \sigma v \right\rangle_{SS\rightarrow A\bar A}$ derived from solutions of Eq.~(\ref{3}). Dark matter masses $m_k,~k=S,A$ are in GeV, $\sigma_{k p}$ in cm$^2$ and various annihilation cross-sections are in cm$^3$ s$^{-1}$ unit.}
        \label{fig1}
    \end{center}
\end{figure}

\begin{figure}
    \begin{center}
        \includegraphics[width=0.5\textwidth]{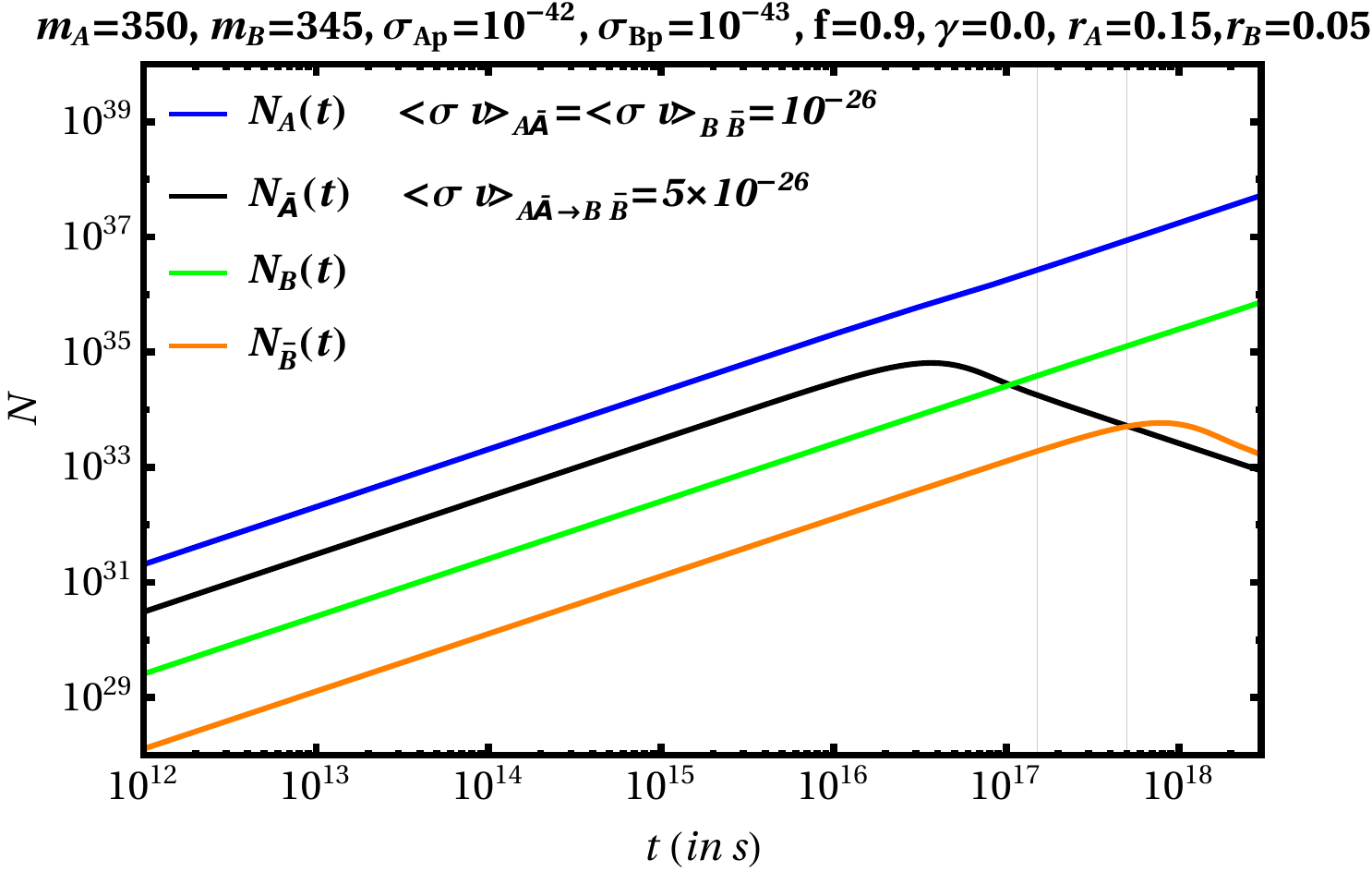}
        \includegraphics[width=0.5\textwidth]{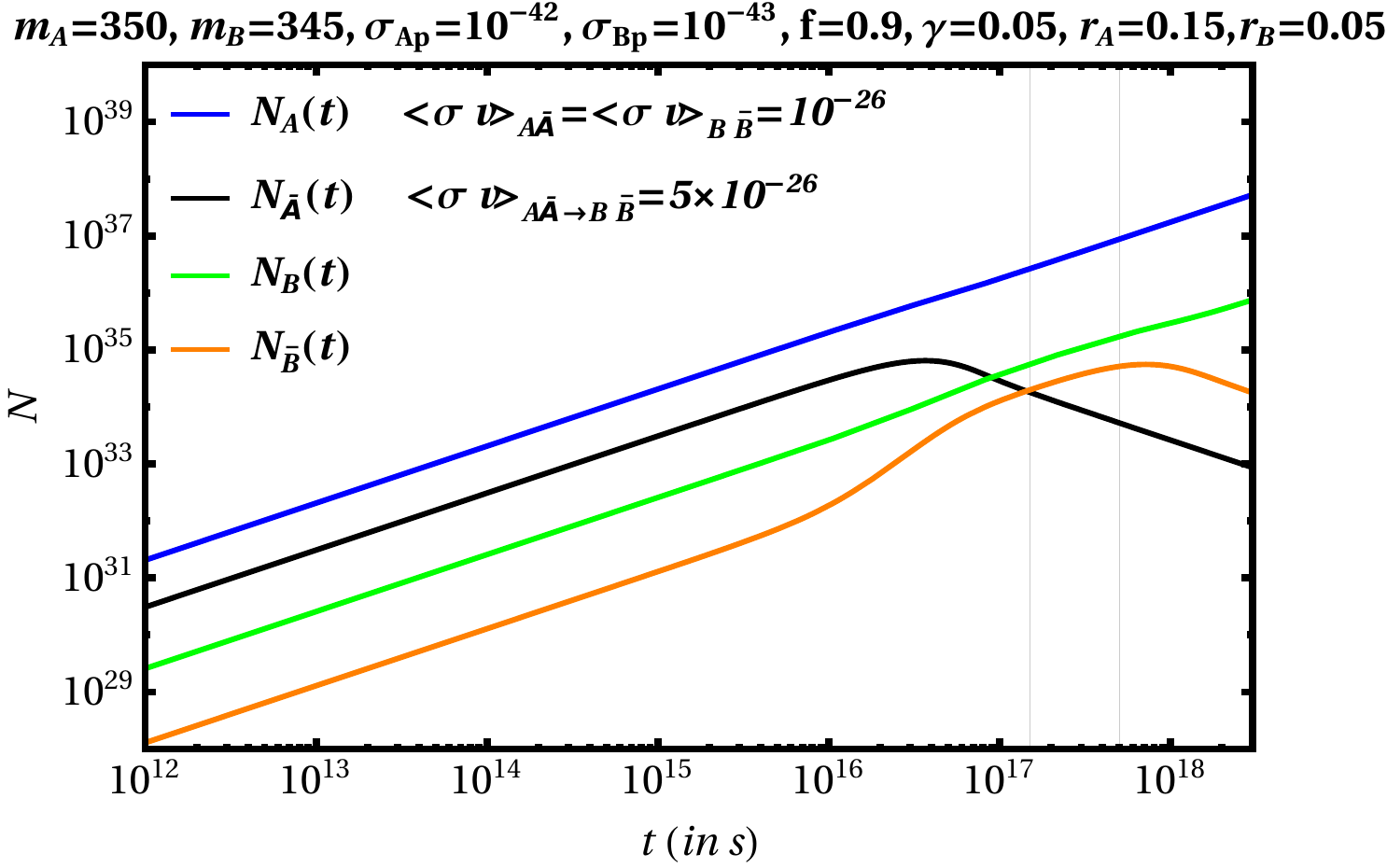}\\
        \includegraphics[width=0.5\textwidth]{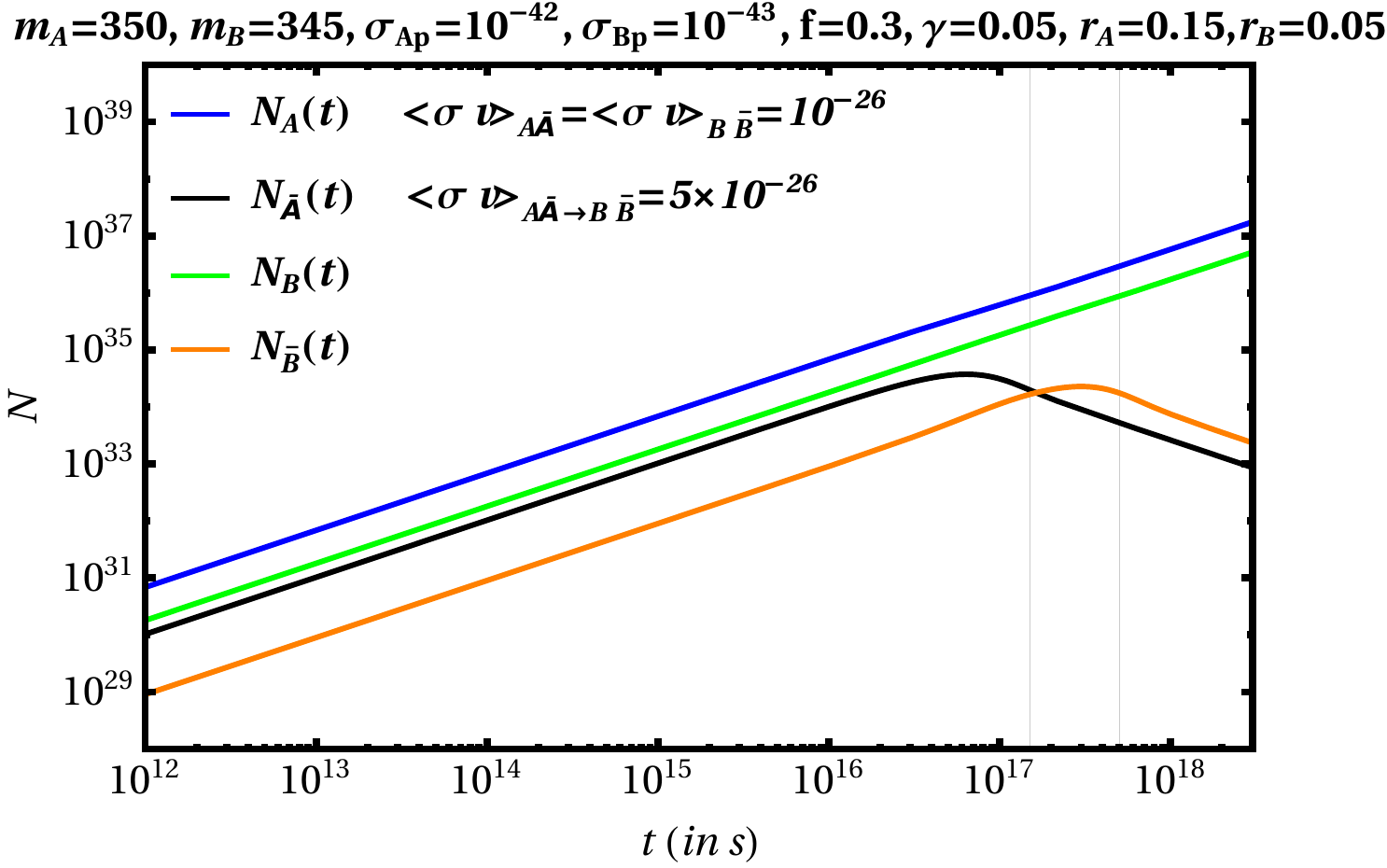}
        \includegraphics[width=0.5\textwidth]{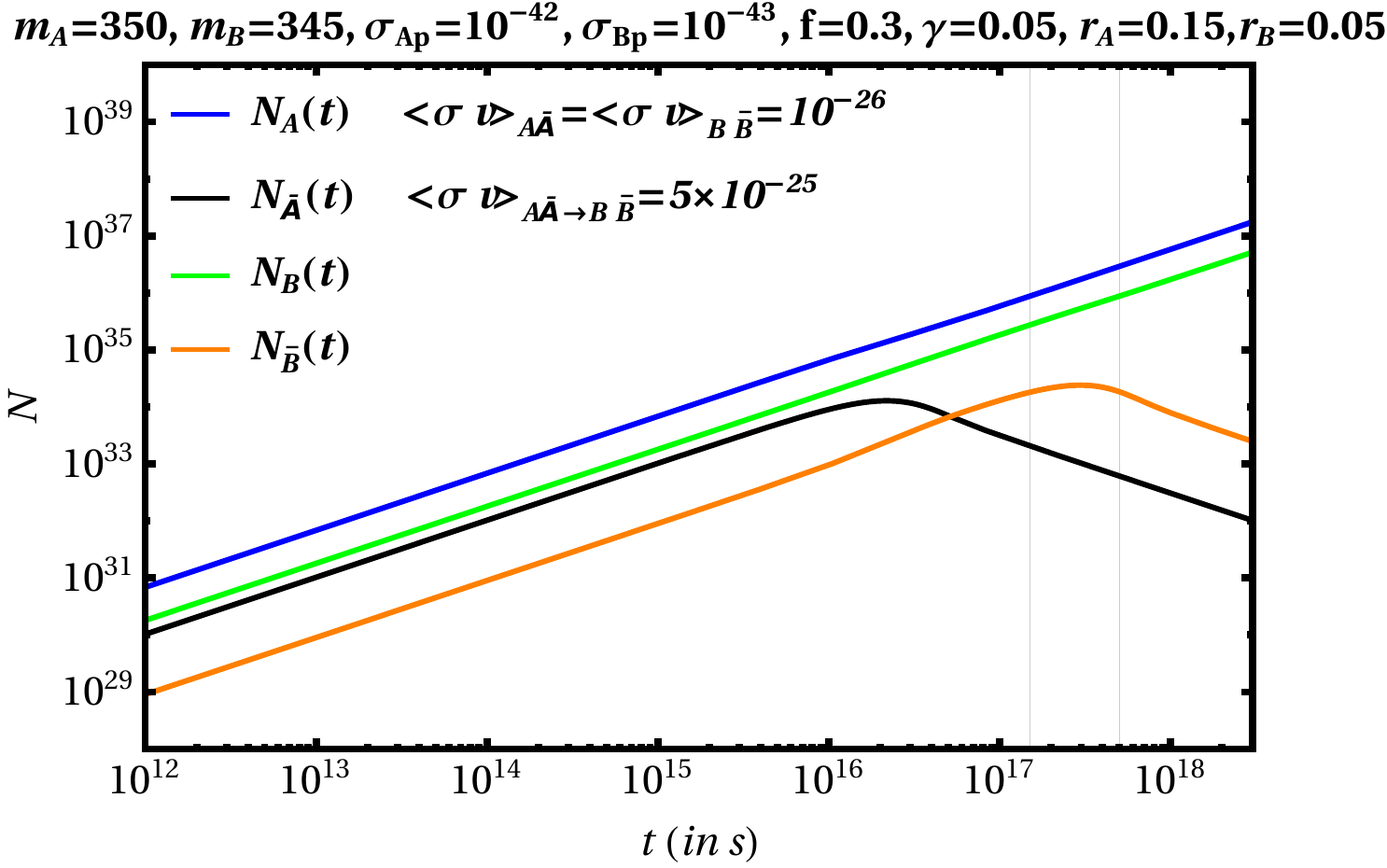}
            \caption{\it Same as Fig.~\ref{fig1} plotted for twin asymmetric dark matter showing $N_{A,\bar A},~N_{B,\bar B}$ variation with time for changes in parameters $\gamma$, $f$, and $\left\langle \sigma v \right\rangle_{A\bar A\rightarrow B\bar B}$. Solutions to $N_{A,\bar A},~N_{B,\bar B}$ is obtained by solving for Eq.~(\ref{4}).}
        \label{fig2}
    \end{center}
\end{figure}
 
\noindent The number of dark matter accumulated inside the Sun can be obtained by solving Eq.~(\ref{3}) for two-component symmetric and asymmetric dark matter systems and Eq.~(\ref{4}) for two asymmetric dark matter frameworks. The solution to Eq.~(\ref{3}) depends on ten independent parameters,
\begin{align}
m_{S},~m_{A},~\sigma_{Sp},~\sigma_{Ap},~f,~\gamma,~r,\nonumber \\~\left\langle \sigma v \right\rangle_{SS},~\left\langle \sigma v \right\rangle_{A\bar A},~\left\langle \sigma v \right\rangle_{SS\rightarrow A\bar A}. 
\label{par1}
\end{align}
Similarly, for the solution of Eq.~(\ref{4}), one extra parameter is required for the solution of DM number abundances inside the Sun
\begin{align}
m_{A},~m_{B},~\sigma_{Ap},~\sigma_{Bp},~f,~\gamma,~r_A,~r_B,\nonumber \\~\left\langle \sigma v \right\rangle_{A\bar A},~\left\langle \sigma v \right\rangle_{B\bar B},~\left\langle \sigma v \right\rangle_{A\bar A\rightarrow B\bar B}.  
\label{par2}
\end{align} 
In Fig.~\ref{fig1}, we solve for Eq.~(\ref{3}) and show the variation of $N_S,~N_{A(\bar A)}$ with time for a specific set of parameter. The vertical lines indicate the ages of the Sun and the Universe. The chosen set of parameters (mass and scattering cross-section) is consistent with the observed limits on the DM spin-dependent scattering cross-section bounds from PICO \cite{PICO:2019vsc} and IceCube \cite{IceCube:2016dgk}. The annihilation cross-section of DM candidates into SM are in agreement with the limits from H.E.S.S. \cite{HESS:2016mib}, Fermi-LAT and MAGIC \cite{MAGIC:2016xys}.
From Fig.~\ref{fig1}, the effect of dark sector annihilation is easily observed comparing the case $\gamma=0$ (no absorption) with a non-zero value of $\gamma=0.05$ for the Sun. 
The effect of changes in the parameter $f$, fraction of relic density shared by one of the  DM candidates, is also visible for different values of $f$, as shown in Fig.~\ref{fig1}. The changes in the $N_{A}$ and $N_{\bar A}$ components will also be governed by hidden sector annihilation. With increased annihilation $\left\langle \sigma v \right\rangle_{SS\rightarrow A\bar A}$ one expects significant changes in $N_A$ and $N_{\bar A}$ population. However, large annihilation reduces the overall abundance of $N_S$ considerably, and the effect of the term $\gamma C^{ann}_{SSA\bar A}  N_S^{2}$ gets reduced. This can be easily observed from the comparison of plots in Fig.~\ref{fig1} for $f=0.3$ shown with two choices of $\left\langle \sigma v \right\rangle_{SS\rightarrow A\bar A}$. Therefore, from Fig.~\ref{fig1}, we observe notable changes in the final number abundances of DM candidates accumulated inside the Sun. We will later investigate how these changes can be traced via neutrino observatories.

\noindent Similar to Fig.~\ref{fig1}, in Fig.~\ref{fig2} we show the evolution of DM number abundances with time inside the Sun for two asymmetric dark matter candidates $A$ and $B$ along with their conjugate components $\bar A$ and $\bar B$. We consider the same set of parameters as in Fig.~\ref{fig1} with $r=r_A$ and an additional parameter $r_B$ for demonstration. In Fig.~\ref{fig2}, the changes in the number of dark matter accumulated in the Sun are observed with the absorption term $\gamma$, the fractional abundances of DM candidates $f$, and the dark sector annihilation $\langle \sigma v\rangle_{A\bar A \rightarrow B\bar B}$. The changes in the DM number abundances follow similar characteristics to Fig.~\ref{fig1} but differ in values due to the asymmetric nature of both candidates.  We also notice that large $\left\langle \sigma v \right\rangle_{A\bar A\rightarrow B\bar B}$ reduces the effect of hidden sector annihilation term in the Sun due to depletion in the value $N_{A}N_{\bar A}$, with moderate reduction $N_{\bar A}$ population.
Therefore, in the case of two asymmetric dark matter candidates, we also expect changes in the event rates or fluxes of dark matter in neutrino detectors, which can be significant but different from the symmetric-asymmetric DM combination.

\begin{figure}
    \begin{center}
        \includegraphics[width=0.45\textwidth]{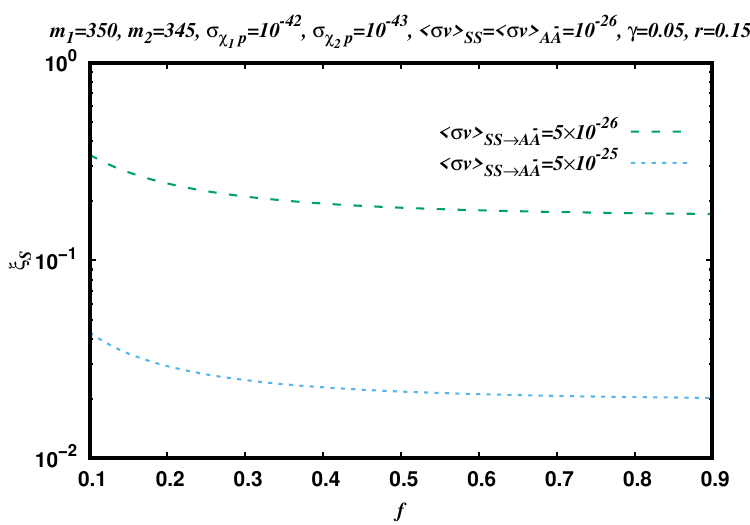}
        \includegraphics[width=0.45\textwidth]{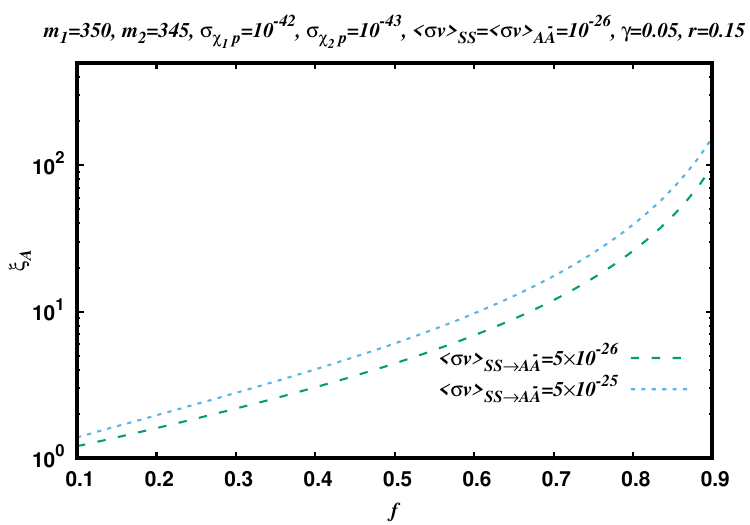}
            \caption{\it Scale factor $\xi_S$ ($\xi_A$) plotted in upper (lower) panel against $f$ for different $\left\langle \sigma v \right\rangle_{SS\rightarrow A\bar A}$ values with fixed set of parameters. Units of various quantities $m_k,~\sigma_{\chi_k}, \left\langle \sigma v \right\rangle_{kk}$, and $\left\langle \sigma v \right\rangle_{SS\rightarrow A\bar A}$ are same as in Fig.~\ref{fig1}.}
        \label{fig3}
    \end{center}
\end{figure}

\begin{figure}
    \begin{center}
        \includegraphics[width=0.45\textwidth]{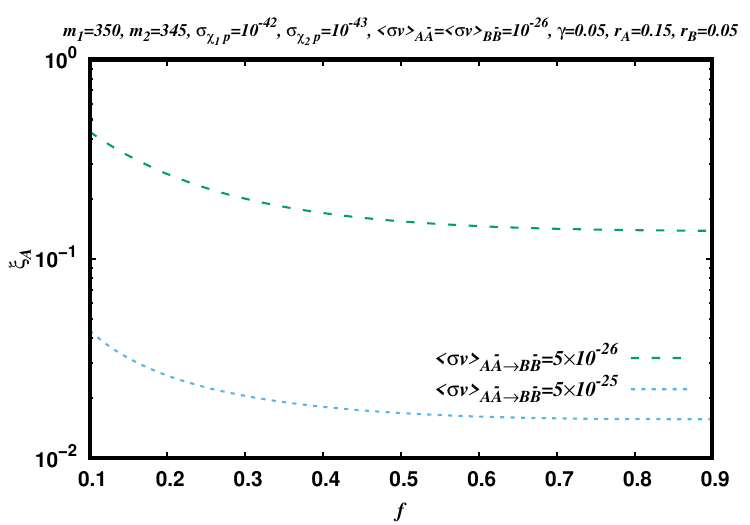}
        \includegraphics[width=0.45\textwidth]{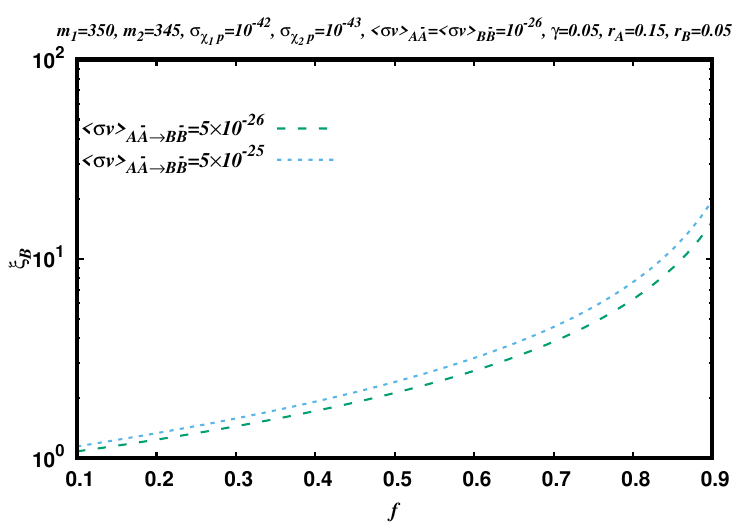}
            \caption{\it Same as Fig.~\ref{fig3} showing variation of scale factors $\xi_A,~\xi_B$ with $f$ and different values of $\left\langle \sigma v \right\rangle_{A\bar A\rightarrow B\bar B}$ for fixed set of parameters.}
        \label{fig4}
    \end{center}
\end{figure}

\noindent We will now discuss how the changes in the number of DM accumulated inside the Sun will contribute to the expected rate of events in neutrino detector experiments. From Fig.~\ref{fig1} and Fig.~\ref{fig2}, we observe that the number of DM accumulated inside the Sun for different species of DMs may not reach equilibrium at $t=t_{\odot}$; $t_{\odot}$ being the age of the Sun. Therefore, a judicious choice is to obtain $N_{k};~k=S, A,\bar A,B,\bar B$ at $t=t_{\odot}$, regardless of the time $t_{\rm eq}$ required by a DM species to reach steady state. With the above consideration, the annihilation flux of various DM species reaching Earth can be expressed as
\begin{align}
\Phi = \frac{\Gamma^{sym/asy}_{\text{ann}}}{4\pi D^2}\, ,
\label{phi}
\end{align}
where 
$\Gamma^{sym}_{\rm{ann}}=\frac{C_{SS}^{ann}}{2} N_S^2(t_\odot)$ for symmetric dark matter and
$\Gamma^{asy}_{\rm{ann}}={C_{k\bar k}^{ann}} N_k(t_\odot) N_{\bar k}(t_\odot)$ for asymmetric dark matter species. The distance between Earth and the Sun is $D=1$ AU or $1.5 \times 10^8$ km.
The corresponding differential muon neutrino flux at Earth is expressed as \cite{Baratella:2013fya}
\begin{equation}
\frac{d\Phi_{\nu_\mu}}{dE_{\nu_\mu}}=\Phi\left(\frac{dN_{\nu_\mu}}{dE_{\nu_\mu}}\right)_{x}\, ,
\label{eq:neutrino_flux}
\end{equation}
with $\frac{dN_{\nu_\mu}}{dE_{\nu_\mu}}$ being the spectrum of $\nu_\mu$ and $\bar{\nu}_\mu$ produced per annihilation of DM candidate into a final state $x$ (lepton, quark, gauge boson etc.). The $\nu_\mu$ ($\bar\nu_\mu$) produces $\mu^-$ ($\mu^+$) particles upon interaction that can be tracked by neutrino detectors. The background of atmospheric muon events can be reduced by tracking up-going muons. The expected muon flux produced by any dark matter species $\chi$ having mass $m_\chi$ is expressed as
\cite{Hooper:2008cf,Erkoca:2009by,Covi:2009xn,Chen:2011vda} 
\begin{eqnarray}
\Phi^{\chi}_{\mu} &=& \int_{E_\mu^{\rm th}}^{m_\chi} d E_\mu \int_{E_\mu}^{m_\chi} d E_{\nu_\mu} \frac{d \Phi_{\nu_\mu}}{d E_{\nu_\mu}}   \times\nonumber \\ 
&& \left[ \frac{\rho}{m_p} \frac{d \sigma_\nu}{d E_\mu} (E_\mu, E_{\nu_\mu}) R_\mu (E_\mu, E_\mu^{\rm th}) \right] + (\nu\to \bar{\nu})\, , \nonumber \\  
\label{fluxmu}
\end{eqnarray}
where
\begin{eqnarray}
R_\mu (E_\mu, E_\mu^{\rm th}) = \frac{1}{\beta \rho} \log \left( \frac{\alpha + \beta E_\mu}{\alpha + \beta E_\mu^{\rm th}} \right)\,,
\label{range}
\end{eqnarray}
is the range of muons with energy $E_{\mu}$ that traverse and lose energy below the threshold energy $E_\mu^{\rm th}$ of the detector. In the Eq.~(\ref{range}) above, 
 $\rho$ denotes water or rock density, $\alpha=2.3\times 10^{-3}\,{\rm cm}^2{\rm g}^{-1} {\rm GeV}^{-1}$ 
and $\beta = 4.4\times 10^{-6}\,{\rm cm}^2{\rm g}^{-1}$. The expression for the neutrino (anti-neutrino) weak interaction scattering cross-section in Eq.~(\ref{fluxmu}) is given as  \cite{Barger:2007xf,Chen:2011vda}
\begin{eqnarray}
\frac{d \sigma^{(p,n)}_\nu(E_\mu, E_{\nu_\mu})}{d E_\mu}  = \frac{2}{\pi} G_F^2 m_p \left( a_\nu^{(p,n)} + b_\nu^{(p,n)} \frac{E_\mu^2}{E_{\nu_\mu}^2}\right)\,, \nonumber \\
\label{nucrs}
\end{eqnarray}
where $a_\nu^{(p,n)}(=b_{\bar \nu}^{(p,n)})=0.15,~0.25$, and $b_\nu^{(p,n)}(=a_{\bar \nu}^{(p,n)})=0.04,~0.06$ for interaction with $\nu$, and $\bar\nu$.
Using the expressions Eq.~(\ref{phi})-(\ref{nucrs}), the expected muon events from DM annihilation inside the Sun can be obtained for a specific detector volume and time of exposure. We introduce a scaling factor $\xi_k;~k=S,A,B$,
defined by the ratio of expected muon events in the presence of the hidden sector annihilation to the case where the evolution of DM candidates is perfectly decoupled, given as
{\small
\bea
\xi_S=\frac{\Phi^S_{\mu}}{\Phi^S_{0\mu}}&=&\frac{\Phi^S}{\Phi^S_0}=\frac{N_S^2(t_\odot)}{N_{S0}^2(t_\odot)}\, ,\nonumber \\
\xi_k=\frac{\Phi^k_{\mu}}{\Phi^k_{0\mu}}&=&\frac{\Phi^k}{\Phi^k_0}=\frac{N_k(t_\odot)N_{\bar k}(t_\odot)}{N_{k0}(t_\odot)N_{\bar k 0}(t_\odot)}\, ;k=A,B\, .
\label{boost}
\eea
}
\noindent In the Eq.~(\ref{boost}) above, suffix zero to the various expressions of muon flux, DM annihilation flux, etc. corresponds to the solution of Eq.~(\ref{3},~\ref{4}) when the hidden sector contributions are set to zero.

\noindent In Fig.~\ref{fig3}, we show the variation of the parameter $\xi_S$
and $\xi_A$ with $f$ and hidden sector annihilation $\left\langle \sigma v \right\rangle_{SS\rightarrow A\bar A}$ keeping other parameters fixed to the values considered in Fig.~\ref{fig1}. With the above choice, capture and visible sector annihilation rates of DM in Eq.~(\ref{3}) being fixed, scaling factor $\xi_{S,A}$ is determined by $f$ and dark sector annihilation only for a particular value of $\gamma$ and $r$. We notice a suppression in the DM flux or events of the symmetric DM candidate, which gradually decreases with high values of $f$. This behaviour can be justified by the changes in the time required to reach steady state directed by $f$ for a chosen $\left\langle \sigma v \right\rangle_{SS\rightarrow A\bar A}$ as we compare with the case where the hidden sector annihilation is absent. We also notice a formidable change in the scaling factor $\xi_S$ for enhancement in dark sector annihilation. A large $\left\langle \sigma v \right\rangle_{SS\rightarrow A\bar A}$ decreases the net $N_S$ population, producing a large number of asymmetric dark matter candidates. However, only a small part of the produced dark matter gets absorbed, determined by the factor $\gamma$. 
On the contrary, we notice that the scale factor $\xi_A$
becomes considerably larger with increasing $f$. This is justified as a large $f$ corresponds to a higher value of $N_S$, contributing to the hidden sector annihilation term $\gamma C^{ann}_{SSA\bar A}  N_S^{2}$. Similarly, for increase in $\left\langle \sigma v \right\rangle_{SS\rightarrow A\bar A}$, we notice an increase in the flux or event rate of asymmetric dark matter for a fixed value of $f$ as the term $C^{ann}_{SSA\bar A}$ becomes large. However, the enhancement is not very prominent,
as large annihilation $\left\langle \sigma v \right\rangle_{SS\rightarrow A\bar A}$
depletes the overall $N_S$ population, countering the effect of $\gamma C^{ann}_{SSA\bar A}  N_S^{2}$ term. Therefore, the number of expected events at neutrino detector experiments for the asymmetric DM candidate gets enhanced considerably when compared with the standard asymmetric DM scenario with no effect of dark sector annihilation. This increases the detection prospects of asymmetric dark matter in neutrino detectors. 

\noindent In Fig.~\ref{fig4}, we replicate the results of Fig.~\ref{fig3}
for the two asymmetric dark matter framework, keeping various parameters fixed at values considered in Fig.~\ref{fig2}. We observe a similar nature in the scaling of $\xi_A$ that resembles
the nature of $\xi_S$ as depicted in Fig.~\ref{fig3}. The suppression in the value $\xi_A$ (or the number of expected events)
is found to be larger with an increase in $f$ when compared to $\xi_S$. This is in part the effect of the fractional asymmetry parameter $r_A$ enforced in the solutions of Eq.~(\ref{4}).
The parameter $\xi_B$ in Fig.~\ref{fig4} is also found to mimic the characteristics of $\xi_A$ shown in Fig.~\ref{fig3} but with a significantly smaller value and almost suppressed by one order of magnitude for large $f$. This outcome is certain as the contribution of the term $\gamma C^{ann}_{A\bar AB \bar B}   N_AN_{\bar A}$ being smaller than $\gamma C^{ann}_{SSA\bar A}  N_S^{2}$ due to the asymmetric nature of DM component $A$ and $\xi_A$ is further reduced by the inclusion of fractional asymmetry parameters $r_A$ and $r_B$. 
From Eq.~(\ref{3}), it is clear that for $\gamma C^{ann}_{SSA\bar A}N_S^{2}\geq \frac{(1-f)r}{1+r}C^c_{\bar A}$,
hidden sector annihilation dominates $N_{\bar A}$ population. The same will happen with $N_{A}$ population inside the Sun
for $\gamma C^{ann}_{SSA\bar A}N_S^{2}\geq \frac{(1-f)}{1+r}C^c_{\bar A}$. Therefore, for small $r$, $N_{\bar A}$ is prominently affected by the hidden sector annihilation. This is also the case for the twin asymmetric dark matter scenario under consideration, and
$N_{\bar B}$ gets modified significantly for $\gamma C^{ann}_{A\bar AB \bar B}   N_AN_{\bar A}\geq \frac{(1-f)r_B}{1+r_B}C^c_{\bar B}$. This explains the nature of the plots shown in Fig.~\ref{fig2}, implying a significant effect on $N_{\bar B}$ but a negligible effect on $N_B$.
However, the overall effect of the hidden sector term $\gamma C^{ann}_{A\bar AB \bar B}   N_AN_{\bar A}$ gets compromised due to the asymmetric property of the dark matter candidate $A$, characterised by the parameter $r_A$, indicating a smaller population of $N_{\bar A}$. We also notice that large hidden sector annihilation $\left\langle \sigma v \right\rangle_{A\bar A\rightarrow B\bar B}$ scales down $N_{\bar A}$ population significantly to reduce the 
effect of $\gamma C^{ann}_{A\bar AB \bar B}   N_AN_{\bar A}$ term.
The smallness of the parameter $\xi_B$ is therefore justified when both DM candidates are asymmetric compared to the parameter $\xi_A$ for the combination of symmetric-asymmetric DM, reflected in the plots of Fig.~\ref{fig3}-\ref{fig4}. Therefore, although an enhancement in the events of the asymmetric DM candidate is observed with two asymmetric dark matter components, the overall signature is less prominent to be registered by detectors.

\section{Conclusions}
\label{con}

\noindent In this work, we propose coupled dark matter frameworks and investigate the dynamics of these coupled dark matter candidates inside the Sun. We consider two possible cases, I) a heavy symmetric dark matter coupled with an asymmetric dark matter, and II) two different asymmetric dark matter candidates coupled with each other.
For both scenarios considered, we notice changes in the dark matter annihilation flux and events to be recorded in neutrino detector experiments based on various parameters of the chosen setup. Beneath, we briefly summarise our findings.
\begin{itemize}
\item For coupled dark matter framework, the hidden sector annihilation of dark matter candidates is responsible for significant changes in the population of DM candidates inside the Sun.
This is observed for both the considerations of an ADM production mechanism inside the Sun, coupled with symmetric and asymmetric dark matter candidates. This observation is based on the comparison between the first two adjacent plots of Fig.~\ref{fig1} for symmetric DM coupled to an asymmetric DM. The same conclusion can be drawn from the similar set of graphs shown in Fig.~\ref{fig2} for two asymmetric DM candidates under consideration.
Effects of hidden sector annihilation becomes prominent when the term $\gamma C^{ann}_{SSA\bar A}  N_S^{2}$
or the term $\gamma C^{ann}_{A\bar AB \bar B}   N_AN_{\bar A}$ dominates over the capture rate term of the DM species.
\item Comparing the second and third plots of Fig.~\ref{fig1}, we notice that the number evolution of DM candidates depends on the parameter $f$, i.e., fractional contribution of a DM to total DM relic abundance. In the present framework, large $f$ indicates a considerable increment in the population of lighter ADM candidates. This behaviour is also observed in the plots of Fig.~\ref{fig2}. A large $f$ corresponds to larger values of $N_S^2$ or
$N_AN_{\bar A}$ resulting in significant production of the ADM candidate while the capture rate of ADM scaled down by a factor $(1-f)$, and, therefore the effects of hidden sector annihilation become prominent.
\item A comparison between the third and fourth adjacent plots of Fig.~\ref{fig1} and Fig.~\ref{fig2} reveals that hidden sector annihilation also plays a key role in determining the accumulated number of a particular DM species. Although a significant increase in the population of lighter species of ADM is expected, an overall reduction in the population of heavier DM $N_S$ or the value of $N_S^2$ and $N_{A,\bar A}$ or the value $N_A N_{\bar A}$ counters the effect.
\item We introduce a scaling factor $\xi_S,~\xi_A,~\xi_B$ to demonstrate the deviations from the expected muon flux (or event rate) obtained from a specific DM defined by the ratio of flux in the presence of hidden sector annihilation to the flux when the hidden sector annihilation term is withdrawn. 
The plots shown in Fig.~\ref{fig3} and Fig.~\ref{fig4} indicate that a sufficient enhancement in the events of low mass ADM can be observed in the present formalism, increasing its detection prospects. However, for the two asymmetric dark matter framework, the factor $\xi_B$ is significantly smaller for large $f$, almost by one order compared to the symmetric-asymmetric framework $\xi_A$.
This decline in the number of expected events is the effect of the twin fractional asymmetry parameters $r_A$ and $r_B$. The enhancement factor or scale factors are only comparable for small $f$ values, when the contribution of the hidden sector annihilation becomes insignificant.
\item For both symmetric-asymmetric and asymmetric-asymmetric DM frameworks, we find a considerable reduction in the annihilation flux (or expected event rate) for the heavier DM candidate as observed in Fig.~\ref{fig3} and Fig.~\ref{fig4}. This reduces the detection prospects of heavier DM candidates coupled to the lighter ADM candidate.
\end{itemize}

\noindent The present work develops the concept and understanding of coupled dark matter dynamics inside the Sun with a focus on asymmetric dark matter. To provide the certainty of dark sector annihilation, we have demonstrated our results on the basis of the solar capture and annihilation of DM candidates with a relevant set of parameters. We do not necessarily intend to solve for coupled Boltzmann equations, which requires a complete model-based approach. For a pure model based analysis, although the final results like scaling factors may differ, the behavioural pattern of results for changes in specific parameters such as $f$, hidden sector annihilations etc., may fairly abide by the findings of the present work. With substantial upgrade to the neutrino detectors, the asymmetric nature of dark matter can be revealed in the future. In the present work, we have only explored multiparticle dark matter and their asymmetry property (if present) in the context of the Sun. Further investigation of coupled ADM can be tested with other astrophysical objects in future studies to bring forth enlightenment towards the understanding of dark matter. 

\noindent {\bf Acknowledgments}: ADB acknowledges financial support from DST, India, under grant number IFA20-PH250 (INSPIRE Faculty Award).

\bibliographystyle{apsrev}
\bibliography{reference}

\begin{thebibliography}{84}
\expandafter\ifx\csname natexlab\endcsname\relax\def\natexlab#1{#1}\fi
\expandafter\ifx\csname bibnamefont\endcsname\relax
  \def\bibnamefont#1{#1}\fi
\expandafter\ifx\csname bibfnamefont\endcsname\relax
  \def\bibfnamefont#1{#1}\fi
\expandafter\ifx\csname citenamefont\endcsname\relax
  \def\citenamefont#1{#1}\fi
\expandafter\ifx\csname url\endcsname\relax
  \def\url#1{\texttt{#1}}\fi
\expandafter\ifx\csname urlprefix\endcsname\relax\def\urlprefix{URL }\fi
\providecommand{\bibinfo}[2]{#2}
\providecommand{\eprint}[2][]{\url{#2}}

\bibitem[{\citenamefont{Aghanim et~al.}(2018)}]{Aghanim:2018eyx}
\bibinfo{author}{\bibfnamefont{N.}~\bibnamefont{Aghanim}} \bibnamefont{et~al.}
  (\bibinfo{collaboration}{Planck}) (\bibinfo{year}{2018}),
  \eprint{1807.06209}.

\bibitem[{\citenamefont{Jungman et~al.}(1996)\citenamefont{Jungman,
  Kamionkowski, and Griest}}]{Jungman:1995df}
\bibinfo{author}{\bibfnamefont{G.}~\bibnamefont{Jungman}},
  \bibinfo{author}{\bibfnamefont{M.}~\bibnamefont{Kamionkowski}},
  \bibnamefont{and} \bibinfo{author}{\bibfnamefont{K.}~\bibnamefont{Griest}},
  \bibinfo{journal}{Phys. Rept.} \textbf{\bibinfo{volume}{267}},
  \bibinfo{pages}{195} (\bibinfo{year}{1996}), \eprint{hep-ph/9506380}.

\bibitem[{\citenamefont{Bertone et~al.}(2005)\citenamefont{Bertone, Hooper, and
  Silk}}]{Bertone:2004pz}
\bibinfo{author}{\bibfnamefont{G.}~\bibnamefont{Bertone}},
  \bibinfo{author}{\bibfnamefont{D.}~\bibnamefont{Hooper}}, \bibnamefont{and}
  \bibinfo{author}{\bibfnamefont{J.}~\bibnamefont{Silk}},
  \bibinfo{journal}{Phys. Rept.} \textbf{\bibinfo{volume}{405}},
  \bibinfo{pages}{279} (\bibinfo{year}{2005}), \eprint{hep-ph/0404175}.

\bibitem[{\citenamefont{Aprile et~al.}(2018)}]{Aprile:2018dbl}
\bibinfo{author}{\bibfnamefont{E.}~\bibnamefont{Aprile}} \bibnamefont{et~al.}
  (\bibinfo{collaboration}{XENON}) (\bibinfo{year}{2018}), \eprint{1805.12562}.

\bibitem[{\citenamefont{Aprile et~al.}(2016)}]{Aprile:2015uzo}
\bibinfo{author}{\bibfnamefont{E.}~\bibnamefont{Aprile}} \bibnamefont{et~al.}
  (\bibinfo{collaboration}{XENON}), \bibinfo{journal}{JCAP}
  \textbf{\bibinfo{volume}{1604}}, \bibinfo{pages}{027} (\bibinfo{year}{2016}),
  \eprint{1512.07501}.

\bibitem[{\citenamefont{Aprile et~al.}(2020)}]{XENON:2020kmp}
\bibinfo{author}{\bibfnamefont{E.}~\bibnamefont{Aprile}} \bibnamefont{et~al.}
  (\bibinfo{collaboration}{XENON}), \bibinfo{journal}{JCAP}
  \textbf{\bibinfo{volume}{11}}, \bibinfo{pages}{031} (\bibinfo{year}{2020}),
  \eprint{2007.08796}.

\bibitem[{\citenamefont{Aprile et~al.}(2023)}]{XENON:2023sxq}
\bibinfo{author}{\bibfnamefont{E.}~\bibnamefont{Aprile}} \bibnamefont{et~al.}
  (\bibinfo{collaboration}{XENON}) (\bibinfo{year}{2023}), \eprint{2303.14729}.

\bibitem[{\citenamefont{Cui et~al.}(2017)}]{Cui:2017nnn}
\bibinfo{author}{\bibfnamefont{X.}~\bibnamefont{Cui}} \bibnamefont{et~al.}
  (\bibinfo{collaboration}{PandaX-II}) (\bibinfo{year}{2017}),
  \eprint{1708.06917}.

\bibitem[{\citenamefont{Amole et~al.}(2019)}]{PICO:2019vsc}
\bibinfo{author}{\bibfnamefont{C.}~\bibnamefont{Amole}} \bibnamefont{et~al.}
  (\bibinfo{collaboration}{PICO}), \bibinfo{journal}{Phys. Rev. D}
  \textbf{\bibinfo{volume}{100}}, \bibinfo{pages}{022001}
  (\bibinfo{year}{2019}), \eprint{1902.04031}.

\bibitem[{\citenamefont{Ackermann et~al.}(2015)}]{Fermi-LAT:2015att}
\bibinfo{author}{\bibfnamefont{M.}~\bibnamefont{Ackermann}}
  \bibnamefont{et~al.} (\bibinfo{collaboration}{Fermi-LAT}),
  \bibinfo{journal}{Phys. Rev. Lett.} \textbf{\bibinfo{volume}{115}},
  \bibinfo{pages}{231301} (\bibinfo{year}{2015}), \eprint{1503.02641}.

\bibitem[{\citenamefont{Drlica-Wagner et~al.}(2015)}]{Fermi-LAT:2015ycq}
\bibinfo{author}{\bibfnamefont{A.}~\bibnamefont{Drlica-Wagner}}
  \bibnamefont{et~al.} (\bibinfo{collaboration}{Fermi-LAT, DES}),
  \bibinfo{journal}{Astrophys. J. Lett.} \textbf{\bibinfo{volume}{809}},
  \bibinfo{pages}{L4} (\bibinfo{year}{2015}), \eprint{1503.02632}.

\bibitem[{\citenamefont{Abdallah et~al.}(2016)}]{HESS:2016mib}
\bibinfo{author}{\bibfnamefont{H.}~\bibnamefont{Abdallah}} \bibnamefont{et~al.}
  (\bibinfo{collaboration}{H.E.S.S.}), \bibinfo{journal}{Phys. Rev. Lett.}
  \textbf{\bibinfo{volume}{117}}, \bibinfo{pages}{111301}
  (\bibinfo{year}{2016}), \eprint{1607.08142}.

\bibitem[{\citenamefont{Ahnen et~al.}(2016)}]{MAGIC:2016xys}
\bibinfo{author}{\bibfnamefont{M.~L.} \bibnamefont{Ahnen}} \bibnamefont{et~al.}
  (\bibinfo{collaboration}{MAGIC, Fermi-LAT}), \bibinfo{journal}{JCAP}
  \textbf{\bibinfo{volume}{02}}, \bibinfo{pages}{039} (\bibinfo{year}{2016}),
  \eprint{1601.06590}.

\bibitem[{\citenamefont{Aartsen et~al.}(2017)}]{IceCube:2016dgk}
\bibinfo{author}{\bibfnamefont{M.~G.} \bibnamefont{Aartsen}}
  \bibnamefont{et~al.} (\bibinfo{collaboration}{IceCube}),
  \bibinfo{journal}{Eur. Phys. J. C} \textbf{\bibinfo{volume}{77}},
  \bibinfo{pages}{146} (\bibinfo{year}{2017}), \bibinfo{note}{[Erratum:
  Eur.Phys.J.C 79, 214 (2019)]}, \eprint{1612.05949}.

\bibitem[{\citenamefont{Choi et~al.}(2015)}]{Super-Kamiokande:2015xms}
\bibinfo{author}{\bibfnamefont{K.}~\bibnamefont{Choi}} \bibnamefont{et~al.}
  (\bibinfo{collaboration}{Super-Kamiokande}), \bibinfo{journal}{Phys. Rev.
  Lett.} \textbf{\bibinfo{volume}{114}}, \bibinfo{pages}{141301}
  (\bibinfo{year}{2015}), \eprint{1503.04858}.

\bibitem[{\citenamefont{Adrian-Martinez et~al.}(2016)}]{ANTARES:2016xuh}
\bibinfo{author}{\bibfnamefont{S.}~\bibnamefont{Adrian-Martinez}}
  \bibnamefont{et~al.} (\bibinfo{collaboration}{ANTARES}),
  \bibinfo{journal}{Phys. Lett. B} \textbf{\bibinfo{volume}{759}},
  \bibinfo{pages}{69} (\bibinfo{year}{2016}), \eprint{1603.02228}.

\bibitem[{\citenamefont{Faulkner and Gilliland}(1985)}]{Faulkner:1985rm}
\bibinfo{author}{\bibfnamefont{J.}~\bibnamefont{Faulkner}} \bibnamefont{and}
  \bibinfo{author}{\bibfnamefont{R.~L.} \bibnamefont{Gilliland}},
  \bibinfo{journal}{Astrophys. J.} \textbf{\bibinfo{volume}{299}},
  \bibinfo{pages}{994} (\bibinfo{year}{1985}).

\bibitem[{\citenamefont{Griest and Seckel}(1987)}]{Griest:1986yu}
\bibinfo{author}{\bibfnamefont{K.}~\bibnamefont{Griest}} \bibnamefont{and}
  \bibinfo{author}{\bibfnamefont{D.}~\bibnamefont{Seckel}},
  \bibinfo{journal}{Nucl. Phys. B} \textbf{\bibinfo{volume}{283}},
  \bibinfo{pages}{681} (\bibinfo{year}{1987}), \bibinfo{note}{[Erratum:
  Nucl.Phys.B 296, 1034--1036 (1988)]}.

\bibitem[{\citenamefont{Gould}(1987)}]{Gould:1987ju}
\bibinfo{author}{\bibfnamefont{A.}~\bibnamefont{Gould}},
  \bibinfo{journal}{Astrophys. J.} \textbf{\bibinfo{volume}{321}},
  \bibinfo{pages}{560} (\bibinfo{year}{1987}).

\bibitem[{\citenamefont{Gould}(1992)}]{Gould:1991hx}
\bibinfo{author}{\bibfnamefont{A.}~\bibnamefont{Gould}},
  \bibinfo{journal}{Astrophys. J.} \textbf{\bibinfo{volume}{388}},
  \bibinfo{pages}{338} (\bibinfo{year}{1992}).

\bibitem[{\citenamefont{Barger et~al.}(2007)\citenamefont{Barger, Keung,
  Shaughnessy, and Tregre}}]{Barger:2007xf}
\bibinfo{author}{\bibfnamefont{V.}~\bibnamefont{Barger}},
  \bibinfo{author}{\bibfnamefont{W.-Y.} \bibnamefont{Keung}},
  \bibinfo{author}{\bibfnamefont{G.}~\bibnamefont{Shaughnessy}},
  \bibnamefont{and} \bibinfo{author}{\bibfnamefont{A.}~\bibnamefont{Tregre}},
  \bibinfo{journal}{Phys. Rev. D} \textbf{\bibinfo{volume}{76}},
  \bibinfo{pages}{095008} (\bibinfo{year}{2007}), \eprint{0708.1325}.

\bibitem[{\citenamefont{Hooper et~al.}(2009)\citenamefont{Hooper, Petriello,
  Zurek, and Kamionkowski}}]{Hooper:2008cf}
\bibinfo{author}{\bibfnamefont{D.}~\bibnamefont{Hooper}},
  \bibinfo{author}{\bibfnamefont{F.}~\bibnamefont{Petriello}},
  \bibinfo{author}{\bibfnamefont{K.~M.} \bibnamefont{Zurek}}, \bibnamefont{and}
  \bibinfo{author}{\bibfnamefont{M.}~\bibnamefont{Kamionkowski}},
  \bibinfo{journal}{Phys. Rev. D} \textbf{\bibinfo{volume}{79}},
  \bibinfo{pages}{015010} (\bibinfo{year}{2009}), \eprint{0808.2464}.

\bibitem[{\citenamefont{Belotsky et~al.}(2009)\citenamefont{Belotsky, Khlopov,
  and Kouvaris}}]{Belotsky:2008vh}
\bibinfo{author}{\bibfnamefont{K.}~\bibnamefont{Belotsky}},
  \bibinfo{author}{\bibfnamefont{M.}~\bibnamefont{Khlopov}}, \bibnamefont{and}
  \bibinfo{author}{\bibfnamefont{C.}~\bibnamefont{Kouvaris}},
  \bibinfo{journal}{Phys. Rev. D} \textbf{\bibinfo{volume}{79}},
  \bibinfo{pages}{083520} (\bibinfo{year}{2009}), \eprint{0810.2022}.

\bibitem[{\citenamefont{Wikstrom and Edsjo}(2009)}]{Wikstrom:2009kw}
\bibinfo{author}{\bibfnamefont{G.}~\bibnamefont{Wikstrom}} \bibnamefont{and}
  \bibinfo{author}{\bibfnamefont{J.}~\bibnamefont{Edsjo}},
  \bibinfo{journal}{JCAP} \textbf{\bibinfo{volume}{04}}, \bibinfo{pages}{009}
  (\bibinfo{year}{2009}), \eprint{0903.2986}.

\bibitem[{\citenamefont{Erkoca et~al.}(2009)\citenamefont{Erkoca, Reno, and
  Sarcevic}}]{Erkoca:2009by}
\bibinfo{author}{\bibfnamefont{A.~E.} \bibnamefont{Erkoca}},
  \bibinfo{author}{\bibfnamefont{M.~H.} \bibnamefont{Reno}}, \bibnamefont{and}
  \bibinfo{author}{\bibfnamefont{I.}~\bibnamefont{Sarcevic}},
  \bibinfo{journal}{Phys. Rev. D} \textbf{\bibinfo{volume}{80}},
  \bibinfo{pages}{043514} (\bibinfo{year}{2009}), \eprint{0906.4364}.

\bibitem[{\citenamefont{Zentner}(2009)}]{Zentner:2009is}
\bibinfo{author}{\bibfnamefont{A.~R.} \bibnamefont{Zentner}},
  \bibinfo{journal}{Phys. Rev. D} \textbf{\bibinfo{volume}{80}},
  \bibinfo{pages}{063501} (\bibinfo{year}{2009}), \eprint{0907.3448}.

\bibitem[{\citenamefont{Covi et~al.}(2010)\citenamefont{Covi, Grefe, Ibarra,
  and Tran}}]{Covi:2009xn}
\bibinfo{author}{\bibfnamefont{L.}~\bibnamefont{Covi}},
  \bibinfo{author}{\bibfnamefont{M.}~\bibnamefont{Grefe}},
  \bibinfo{author}{\bibfnamefont{A.}~\bibnamefont{Ibarra}}, \bibnamefont{and}
  \bibinfo{author}{\bibfnamefont{D.}~\bibnamefont{Tran}},
  \bibinfo{journal}{JCAP} \textbf{\bibinfo{volume}{04}}, \bibinfo{pages}{017}
  (\bibinfo{year}{2010}), \eprint{0912.3521}.

\bibitem[{\citenamefont{Chen and Zhang}(2011)}]{Chen:2011vda}
\bibinfo{author}{\bibfnamefont{S.-L.} \bibnamefont{Chen}} \bibnamefont{and}
  \bibinfo{author}{\bibfnamefont{Y.}~\bibnamefont{Zhang}},
  \bibinfo{journal}{Phys. Rev. D} \textbf{\bibinfo{volume}{84}},
  \bibinfo{pages}{031301} (\bibinfo{year}{2011}), \eprint{1106.4044}.

\bibitem[{\citenamefont{Bernal et~al.}(2013)\citenamefont{Bernal,
  Mart\'\i{}n-Albo, and Palomares-Ruiz}}]{Bernal:2012qh}
\bibinfo{author}{\bibfnamefont{N.}~\bibnamefont{Bernal}},
  \bibinfo{author}{\bibfnamefont{J.}~\bibnamefont{Mart\'\i{}n-Albo}},
  \bibnamefont{and}
  \bibinfo{author}{\bibfnamefont{S.}~\bibnamefont{Palomares-Ruiz}},
  \bibinfo{journal}{JCAP} \textbf{\bibinfo{volume}{08}}, \bibinfo{pages}{011}
  (\bibinfo{year}{2013}), \eprint{1208.0834}.

\bibitem[{\citenamefont{Chen et~al.}(2014)\citenamefont{Chen, Lee, Lin, and
  Lin}}]{Chen:2014oaa}
\bibinfo{author}{\bibfnamefont{C.-S.} \bibnamefont{Chen}},
  \bibinfo{author}{\bibfnamefont{F.-F.} \bibnamefont{Lee}},
  \bibinfo{author}{\bibfnamefont{G.-L.} \bibnamefont{Lin}}, \bibnamefont{and}
  \bibinfo{author}{\bibfnamefont{Y.-H.} \bibnamefont{Lin}},
  \bibinfo{journal}{JCAP} \textbf{\bibinfo{volume}{10}}, \bibinfo{pages}{049}
  (\bibinfo{year}{2014}), \eprint{1408.5471}.

\bibitem[{\citenamefont{Catena and Widmark}(2016)}]{Catena:2016ckl}
\bibinfo{author}{\bibfnamefont{R.}~\bibnamefont{Catena}} \bibnamefont{and}
  \bibinfo{author}{\bibfnamefont{A.}~\bibnamefont{Widmark}},
  \bibinfo{journal}{JCAP} \textbf{\bibinfo{volume}{12}}, \bibinfo{pages}{016}
  (\bibinfo{year}{2016}), \eprint{1609.04825}.

\bibitem[{\citenamefont{Tiwari et~al.}(2019)\citenamefont{Tiwari, Choubey, and
  Ghosh}}]{Tiwari:2018gxz}
\bibinfo{author}{\bibfnamefont{D.}~\bibnamefont{Tiwari}},
  \bibinfo{author}{\bibfnamefont{S.}~\bibnamefont{Choubey}}, \bibnamefont{and}
  \bibinfo{author}{\bibfnamefont{A.}~\bibnamefont{Ghosh}},
  \bibinfo{journal}{JHEP} \textbf{\bibinfo{volume}{05}}, \bibinfo{pages}{039}
  (\bibinfo{year}{2019}), \eprint{1806.05058}.

\bibitem[{\citenamefont{Gaidau and Shelton}(2019)}]{Gaidau:2018yws}
\bibinfo{author}{\bibfnamefont{C.}~\bibnamefont{Gaidau}} \bibnamefont{and}
  \bibinfo{author}{\bibfnamefont{J.}~\bibnamefont{Shelton}},
  \bibinfo{journal}{JCAP} \textbf{\bibinfo{volume}{06}}, \bibinfo{pages}{022}
  (\bibinfo{year}{2019}), \eprint{1811.00557}.

\bibitem[{\citenamefont{Gupta et~al.}(2022)\citenamefont{Gupta, Majumdar, and
  Halder}}]{Gupta:2022lws}
\bibinfo{author}{\bibfnamefont{A.}~\bibnamefont{Gupta}},
  \bibinfo{author}{\bibfnamefont{D.}~\bibnamefont{Majumdar}}, \bibnamefont{and}
  \bibinfo{author}{\bibfnamefont{A.}~\bibnamefont{Halder}},
  \bibinfo{journal}{Mod. Phys. Lett. A} \textbf{\bibinfo{volume}{37}},
  \bibinfo{pages}{2250233} (\bibinfo{year}{2022}), \eprint{2203.13697}.

\bibitem[{\citenamefont{Belanger and Park}(2012)}]{Belanger:2011ww}
\bibinfo{author}{\bibfnamefont{G.}~\bibnamefont{Belanger}} \bibnamefont{and}
  \bibinfo{author}{\bibfnamefont{J.-C.} \bibnamefont{Park}},
  \bibinfo{journal}{JCAP} \textbf{\bibinfo{volume}{03}}, \bibinfo{pages}{038}
  (\bibinfo{year}{2012}), \eprint{1112.4491}.

\bibitem[{\citenamefont{Khlopov}(1995)}]{Khlopov:1995pa}
\bibinfo{author}{\bibfnamefont{M.~Y.} \bibnamefont{Khlopov}}, in
  \emph{\bibinfo{booktitle}{{30th Rencontres de Moriond: Euroconferences: Dark
  Matter in Cosmology, Clocks and Tests of Fundamental Laws}}}
  (\bibinfo{year}{1995}), pp. \bibinfo{pages}{133--138}.

\bibitem[{\citenamefont{Bhattacharya et~al.}(2013)\citenamefont{Bhattacharya,
  Drozd, Grzadkowski, and Wudka}}]{Bhattacharya:2013hva}
\bibinfo{author}{\bibfnamefont{S.}~\bibnamefont{Bhattacharya}},
  \bibinfo{author}{\bibfnamefont{A.}~\bibnamefont{Drozd}},
  \bibinfo{author}{\bibfnamefont{B.}~\bibnamefont{Grzadkowski}},
  \bibnamefont{and} \bibinfo{author}{\bibfnamefont{J.}~\bibnamefont{Wudka}},
  \bibinfo{journal}{JHEP} \textbf{\bibinfo{volume}{10}}, \bibinfo{pages}{158}
  (\bibinfo{year}{2013}), \eprint{1309.2986}.

\bibitem[{\citenamefont{Bian et~al.}(2014)\citenamefont{Bian, Ding, and
  Zhu}}]{Bian:2013wna}
\bibinfo{author}{\bibfnamefont{L.}~\bibnamefont{Bian}},
  \bibinfo{author}{\bibfnamefont{R.}~\bibnamefont{Ding}}, \bibnamefont{and}
  \bibinfo{author}{\bibfnamefont{B.}~\bibnamefont{Zhu}},
  \bibinfo{journal}{Phys. Lett.} \textbf{\bibinfo{volume}{B728}},
  \bibinfo{pages}{105} (\bibinfo{year}{2014}), \eprint{1308.3851}.

\bibitem[{\citenamefont{Esch et~al.}(2014)\citenamefont{Esch, Klasen, and
  Yaguna}}]{Esch:2014jpa}
\bibinfo{author}{\bibfnamefont{S.}~\bibnamefont{Esch}},
  \bibinfo{author}{\bibfnamefont{M.}~\bibnamefont{Klasen}}, \bibnamefont{and}
  \bibinfo{author}{\bibfnamefont{C.~E.} \bibnamefont{Yaguna}},
  \bibinfo{journal}{JHEP} \textbf{\bibinfo{volume}{09}}, \bibinfo{pages}{108}
  (\bibinfo{year}{2014}), \eprint{1406.0617}.

\bibitem[{\citenamefont{Ahmed et~al.}(2018)\citenamefont{Ahmed, Duch,
  Grzadkowski, and Iglicki}}]{Ahmed:2017dbb}
\bibinfo{author}{\bibfnamefont{A.}~\bibnamefont{Ahmed}},
  \bibinfo{author}{\bibfnamefont{M.}~\bibnamefont{Duch}},
  \bibinfo{author}{\bibfnamefont{B.}~\bibnamefont{Grzadkowski}},
  \bibnamefont{and} \bibinfo{author}{\bibfnamefont{M.}~\bibnamefont{Iglicki}},
  \bibinfo{journal}{Eur. Phys. J.} \textbf{\bibinfo{volume}{C78}},
  \bibinfo{pages}{905} (\bibinfo{year}{2018}), \eprint{1710.01853}.

\bibitem[{\citenamefont{Herrero-Garcia
  et~al.}(2017)\citenamefont{Herrero-Garcia, Scaffidi, White, and
  Williams}}]{Herrero-Garcia:2017vrl}
\bibinfo{author}{\bibfnamefont{J.}~\bibnamefont{Herrero-Garcia}},
  \bibinfo{author}{\bibfnamefont{A.}~\bibnamefont{Scaffidi}},
  \bibinfo{author}{\bibfnamefont{M.}~\bibnamefont{White}}, \bibnamefont{and}
  \bibinfo{author}{\bibfnamefont{A.~G.} \bibnamefont{Williams}},
  \bibinfo{journal}{JCAP} \textbf{\bibinfo{volume}{1711}}, \bibinfo{pages}{021}
  (\bibinfo{year}{2017}), \eprint{1709.01945}.

\bibitem[{\citenamefont{Herrero-Garcia
  et~al.}(2019)\citenamefont{Herrero-Garcia, Scaffidi, White, and
  Williams}}]{Herrero-Garcia:2018qnz}
\bibinfo{author}{\bibfnamefont{J.}~\bibnamefont{Herrero-Garcia}},
  \bibinfo{author}{\bibfnamefont{A.}~\bibnamefont{Scaffidi}},
  \bibinfo{author}{\bibfnamefont{M.}~\bibnamefont{White}}, \bibnamefont{and}
  \bibinfo{author}{\bibfnamefont{A.~G.} \bibnamefont{Williams}},
  \bibinfo{journal}{JCAP} \textbf{\bibinfo{volume}{1901}}, \bibinfo{pages}{008}
  (\bibinfo{year}{2019}), \eprint{1809.06881}.

\bibitem[{\citenamefont{Aoki and Toma}(2018)}]{Aoki:2018gjf}
\bibinfo{author}{\bibfnamefont{M.}~\bibnamefont{Aoki}} \bibnamefont{and}
  \bibinfo{author}{\bibfnamefont{T.}~\bibnamefont{Toma}},
  \bibinfo{journal}{JCAP} \textbf{\bibinfo{volume}{1810}}, \bibinfo{pages}{020}
  (\bibinfo{year}{2018}), \eprint{1806.09154}.

\bibitem[{\citenamefont{Bhattacharya et~al.}(2017)\citenamefont{Bhattacharya,
  Poulose, and Ghosh}}]{Bhattacharya:2016ysw}
\bibinfo{author}{\bibfnamefont{S.}~\bibnamefont{Bhattacharya}},
  \bibinfo{author}{\bibfnamefont{P.}~\bibnamefont{Poulose}}, \bibnamefont{and}
  \bibinfo{author}{\bibfnamefont{P.}~\bibnamefont{Ghosh}},
  \bibinfo{journal}{JCAP} \textbf{\bibinfo{volume}{1704}}, \bibinfo{pages}{043}
  (\bibinfo{year}{2017}), \eprint{1607.08461}.

\bibitem[{\citenamefont{Chakraborti et~al.}(2019)\citenamefont{Chakraborti,
  Dutta~Banik, and Islam}}]{Chakraborti:2018aae}
\bibinfo{author}{\bibfnamefont{S.}~\bibnamefont{Chakraborti}},
  \bibinfo{author}{\bibfnamefont{A.}~\bibnamefont{Dutta~Banik}},
  \bibnamefont{and} \bibinfo{author}{\bibfnamefont{R.}~\bibnamefont{Islam}},
  \bibinfo{journal}{Eur. Phys. J. C} \textbf{\bibinfo{volume}{79}},
  \bibinfo{pages}{662} (\bibinfo{year}{2019}), \eprint{1810.05595}.

\bibitem[{\citenamefont{Elahi and Khatibi}(2019)}]{Elahi:2019jeo}
\bibinfo{author}{\bibfnamefont{F.}~\bibnamefont{Elahi}} \bibnamefont{and}
  \bibinfo{author}{\bibfnamefont{S.}~\bibnamefont{Khatibi}},
  \bibinfo{journal}{Phys. Rev.} \textbf{\bibinfo{volume}{D100}},
  \bibinfo{pages}{015019} (\bibinfo{year}{2019}), \eprint{1902.04384}.

\bibitem[{\citenamefont{Borah et~al.}(2019)\citenamefont{Borah, Roshan, and
  Sil}}]{Borah:2019aeq}
\bibinfo{author}{\bibfnamefont{D.}~\bibnamefont{Borah}},
  \bibinfo{author}{\bibfnamefont{R.}~\bibnamefont{Roshan}}, \bibnamefont{and}
  \bibinfo{author}{\bibfnamefont{A.}~\bibnamefont{Sil}},
  \bibinfo{journal}{Phys. Rev. D} \textbf{\bibinfo{volume}{100}},
  \bibinfo{pages}{055027} (\bibinfo{year}{2019}), \eprint{1904.04837}.

\bibitem[{\citenamefont{Bhattacharya et~al.}(2020)\citenamefont{Bhattacharya,
  Chakrabarty, Roshan, and Sil}}]{Bhattacharya:2019tqq}
\bibinfo{author}{\bibfnamefont{S.}~\bibnamefont{Bhattacharya}},
  \bibinfo{author}{\bibfnamefont{N.}~\bibnamefont{Chakrabarty}},
  \bibinfo{author}{\bibfnamefont{R.}~\bibnamefont{Roshan}}, \bibnamefont{and}
  \bibinfo{author}{\bibfnamefont{A.}~\bibnamefont{Sil}},
  \bibinfo{journal}{JCAP} \textbf{\bibinfo{volume}{04}}, \bibinfo{pages}{013}
  (\bibinfo{year}{2020}), \eprint{1910.00612}.

\bibitem[{\citenamefont{Dutta~Banik et~al.}(2021)\citenamefont{Dutta~Banik,
  Roshan, and Sil}}]{DuttaBanik:2020jrj}
\bibinfo{author}{\bibfnamefont{A.}~\bibnamefont{Dutta~Banik}},
  \bibinfo{author}{\bibfnamefont{R.}~\bibnamefont{Roshan}}, \bibnamefont{and}
  \bibinfo{author}{\bibfnamefont{A.}~\bibnamefont{Sil}},
  \bibinfo{journal}{Phys. Rev. D} \textbf{\bibinfo{volume}{103}},
  \bibinfo{pages}{075001} (\bibinfo{year}{2021}), \eprint{2009.01262}.

\bibitem[{\citenamefont{D\'\i{}az~S\'aez
  et~al.}(2021)\citenamefont{D\'\i{}az~S\'aez, M\"ohling, and
  St\"ockinger}}]{DiazSaez:2021pfw}
\bibinfo{author}{\bibfnamefont{B.}~\bibnamefont{D\'\i{}az~S\'aez}},
  \bibinfo{author}{\bibfnamefont{K.}~\bibnamefont{M\"ohling}},
  \bibnamefont{and}
  \bibinfo{author}{\bibfnamefont{D.}~\bibnamefont{St\"ockinger}},
  \bibinfo{journal}{JCAP} \textbf{\bibinfo{volume}{10}}, \bibinfo{pages}{027}
  (\bibinfo{year}{2021}), \eprint{2103.17064}.

\bibitem[{\citenamefont{Bas~i Beneito et~al.}(2022)\citenamefont{Bas~i Beneito,
  Herrero-Garc\'\i{}a, and Vatsyayan}}]{BasiBeneito:2022qxd}
\bibinfo{author}{\bibfnamefont{A.}~\bibnamefont{Bas~i Beneito}},
  \bibinfo{author}{\bibfnamefont{J.}~\bibnamefont{Herrero-Garc\'\i{}a}},
  \bibnamefont{and}
  \bibinfo{author}{\bibfnamefont{D.}~\bibnamefont{Vatsyayan}},
  \bibinfo{journal}{JHEP} \textbf{\bibinfo{volume}{10}}, \bibinfo{pages}{075}
  (\bibinfo{year}{2022}), \eprint{2207.02874}.

\bibitem[{\citenamefont{Elor and McGehee}(2021)}]{Elor:2020tkc}
\bibinfo{author}{\bibfnamefont{G.}~\bibnamefont{Elor}} \bibnamefont{and}
  \bibinfo{author}{\bibfnamefont{R.}~\bibnamefont{McGehee}},
  \bibinfo{journal}{Phys. Rev. D} \textbf{\bibinfo{volume}{103}},
  \bibinfo{pages}{035005} (\bibinfo{year}{2021}), \eprint{2011.06115}.

\bibitem[{\citenamefont{Elahi et~al.}(2022)\citenamefont{Elahi, Elor, and
  McGehee}}]{Elahi:2021jia}
\bibinfo{author}{\bibfnamefont{F.}~\bibnamefont{Elahi}},
  \bibinfo{author}{\bibfnamefont{G.}~\bibnamefont{Elor}}, \bibnamefont{and}
  \bibinfo{author}{\bibfnamefont{R.}~\bibnamefont{McGehee}},
  \bibinfo{journal}{Phys. Rev. D} \textbf{\bibinfo{volume}{105}},
  \bibinfo{pages}{055024} (\bibinfo{year}{2022}), \eprint{2109.09751}.

\bibitem[{\citenamefont{Aoki et~al.}(2012)\citenamefont{Aoki, Duerr, Kubo, and
  Takano}}]{Aoki:2012ub}
\bibinfo{author}{\bibfnamefont{M.}~\bibnamefont{Aoki}},
  \bibinfo{author}{\bibfnamefont{M.}~\bibnamefont{Duerr}},
  \bibinfo{author}{\bibfnamefont{J.}~\bibnamefont{Kubo}}, \bibnamefont{and}
  \bibinfo{author}{\bibfnamefont{H.}~\bibnamefont{Takano}},
  \bibinfo{journal}{Phys. Rev. D} \textbf{\bibinfo{volume}{86}},
  \bibinfo{pages}{076015} (\bibinfo{year}{2012}), \eprint{1207.3318}.

\bibitem[{\citenamefont{Aoki et~al.}(2014)\citenamefont{Aoki, Kubo, and
  Takano}}]{Aoki:2014lha}
\bibinfo{author}{\bibfnamefont{M.}~\bibnamefont{Aoki}},
  \bibinfo{author}{\bibfnamefont{J.}~\bibnamefont{Kubo}}, \bibnamefont{and}
  \bibinfo{author}{\bibfnamefont{H.}~\bibnamefont{Takano}},
  \bibinfo{journal}{Phys. Rev. D} \textbf{\bibinfo{volume}{90}},
  \bibinfo{pages}{076011} (\bibinfo{year}{2014}), \eprint{1408.1853}.

\bibitem[{\citenamefont{Berger et~al.}(2015)\citenamefont{Berger, Cui, and
  Zhao}}]{Berger:2014sqa}
\bibinfo{author}{\bibfnamefont{J.}~\bibnamefont{Berger}},
  \bibinfo{author}{\bibfnamefont{Y.}~\bibnamefont{Cui}}, \bibnamefont{and}
  \bibinfo{author}{\bibfnamefont{Y.}~\bibnamefont{Zhao}},
  \bibinfo{journal}{JCAP} \textbf{\bibinfo{volume}{02}}, \bibinfo{pages}{005}
  (\bibinfo{year}{2015}), \eprint{1410.2246}.

\bibitem[{\citenamefont{Aoki et~al.}(2017)\citenamefont{Aoki, Kaneko, and
  Kubo}}]{Aoki:2017eqn}
\bibinfo{author}{\bibfnamefont{M.}~\bibnamefont{Aoki}},
  \bibinfo{author}{\bibfnamefont{D.}~\bibnamefont{Kaneko}}, \bibnamefont{and}
  \bibinfo{author}{\bibfnamefont{J.}~\bibnamefont{Kubo}},
  \bibinfo{journal}{Front. in Phys.} \textbf{\bibinfo{volume}{5}},
  \bibinfo{pages}{53} (\bibinfo{year}{2017}), \eprint{1711.03765}.

\bibitem[{\citenamefont{Dutta~Banik}(2024)}]{DuttaBanik:2023yxj}
\bibinfo{author}{\bibfnamefont{A.}~\bibnamefont{Dutta~Banik}},
  \bibinfo{journal}{Nucl. Phys. B} \textbf{\bibinfo{volume}{998}},
  \bibinfo{pages}{116394} (\bibinfo{year}{2024}), \eprint{2304.04721}.

\bibitem[{\citenamefont{Kaplan et~al.}(2009)\citenamefont{Kaplan, Luty, and
  Zurek}}]{Kaplan:2009ag}
\bibinfo{author}{\bibfnamefont{D.~E.} \bibnamefont{Kaplan}},
  \bibinfo{author}{\bibfnamefont{M.~A.} \bibnamefont{Luty}}, \bibnamefont{and}
  \bibinfo{author}{\bibfnamefont{K.~M.} \bibnamefont{Zurek}},
  \bibinfo{journal}{Phys. Rev. D} \textbf{\bibinfo{volume}{79}},
  \bibinfo{pages}{115016} (\bibinfo{year}{2009}), \eprint{0901.4117}.

\bibitem[{\citenamefont{Graesser et~al.}(2011)\citenamefont{Graesser,
  Shoemaker, and Vecchi}}]{Graesser:2011wi}
\bibinfo{author}{\bibfnamefont{M.~L.} \bibnamefont{Graesser}},
  \bibinfo{author}{\bibfnamefont{I.~M.} \bibnamefont{Shoemaker}},
  \bibnamefont{and} \bibinfo{author}{\bibfnamefont{L.}~\bibnamefont{Vecchi}},
  \bibinfo{journal}{JHEP} \textbf{\bibinfo{volume}{10}}, \bibinfo{pages}{110}
  (\bibinfo{year}{2011}), \eprint{1103.2771}.

\bibitem[{\citenamefont{Iminniyaz et~al.}(2011)\citenamefont{Iminniyaz, Drees,
  and Chen}}]{Iminniyaz:2011yp}
\bibinfo{author}{\bibfnamefont{H.}~\bibnamefont{Iminniyaz}},
  \bibinfo{author}{\bibfnamefont{M.}~\bibnamefont{Drees}}, \bibnamefont{and}
  \bibinfo{author}{\bibfnamefont{X.}~\bibnamefont{Chen}},
  \bibinfo{journal}{JCAP} \textbf{\bibinfo{volume}{07}}, \bibinfo{pages}{003}
  (\bibinfo{year}{2011}), \eprint{1104.5548}.

\bibitem[{\citenamefont{Gelmini et~al.}(2013)\citenamefont{Gelmini, Huh, and
  Rehagen}}]{Gelmini:2013awa}
\bibinfo{author}{\bibfnamefont{G.~B.} \bibnamefont{Gelmini}},
  \bibinfo{author}{\bibfnamefont{J.-H.} \bibnamefont{Huh}}, \bibnamefont{and}
  \bibinfo{author}{\bibfnamefont{T.}~\bibnamefont{Rehagen}},
  \bibinfo{journal}{JCAP} \textbf{\bibinfo{volume}{08}}, \bibinfo{pages}{003}
  (\bibinfo{year}{2013}), \eprint{1304.3679}.

\bibitem[{\citenamefont{Zurek}(2014)}]{Zurek:2013wia}
\bibinfo{author}{\bibfnamefont{K.~M.} \bibnamefont{Zurek}},
  \bibinfo{journal}{Phys. Rept.} \textbf{\bibinfo{volume}{537}},
  \bibinfo{pages}{91} (\bibinfo{year}{2014}), \eprint{1308.0338}.

\bibitem[{\citenamefont{Kitabayashi and Kurosawa}(2016)}]{Kitabayashi:2015oda}
\bibinfo{author}{\bibfnamefont{T.}~\bibnamefont{Kitabayashi}} \bibnamefont{and}
  \bibinfo{author}{\bibfnamefont{Y.}~\bibnamefont{Kurosawa}},
  \bibinfo{journal}{Phys. Rev. D} \textbf{\bibinfo{volume}{93}},
  \bibinfo{pages}{033002} (\bibinfo{year}{2016}), \eprint{1509.05564}.

\bibitem[{\citenamefont{Blennow and Clementz}(2015)}]{Blennow:2015xha}
\bibinfo{author}{\bibfnamefont{M.}~\bibnamefont{Blennow}} \bibnamefont{and}
  \bibinfo{author}{\bibfnamefont{S.}~\bibnamefont{Clementz}},
  \bibinfo{journal}{JCAP} \textbf{\bibinfo{volume}{08}}, \bibinfo{pages}{036}
  (\bibinfo{year}{2015}), \eprint{1504.05813}.

\bibitem[{\citenamefont{Iminniyaz}(2017)}]{Iminniyaz:2016iom}
\bibinfo{author}{\bibfnamefont{H.}~\bibnamefont{Iminniyaz}},
  \bibinfo{journal}{Phys. Lett. B} \textbf{\bibinfo{volume}{765}},
  \bibinfo{pages}{6} (\bibinfo{year}{2017}), \eprint{1604.04251}.

\bibitem[{\citenamefont{Agrawal et~al.}(2017)\citenamefont{Agrawal, Kilic,
  Swaminathan, and Trendafilova}}]{Agrawal:2016uwf}
\bibinfo{author}{\bibfnamefont{P.}~\bibnamefont{Agrawal}},
  \bibinfo{author}{\bibfnamefont{C.}~\bibnamefont{Kilic}},
  \bibinfo{author}{\bibfnamefont{S.}~\bibnamefont{Swaminathan}},
  \bibnamefont{and}
  \bibinfo{author}{\bibfnamefont{C.}~\bibnamefont{Trendafilova}},
  \bibinfo{journal}{Phys. Rev. D} \textbf{\bibinfo{volume}{95}},
  \bibinfo{pages}{015031} (\bibinfo{year}{2017}), \eprint{1608.04745}.

\bibitem[{\citenamefont{Nagata et~al.}(2017)\citenamefont{Nagata, Olive, and
  Zheng}}]{Nagata:2016knk}
\bibinfo{author}{\bibfnamefont{N.}~\bibnamefont{Nagata}},
  \bibinfo{author}{\bibfnamefont{K.~A.} \bibnamefont{Olive}}, \bibnamefont{and}
  \bibinfo{author}{\bibfnamefont{J.}~\bibnamefont{Zheng}},
  \bibinfo{journal}{JCAP} \textbf{\bibinfo{volume}{02}}, \bibinfo{pages}{016}
  (\bibinfo{year}{2017}), \eprint{1611.04693}.

\bibitem[{\citenamefont{Baldes and Petraki}(2017)}]{Baldes:2017gzw}
\bibinfo{author}{\bibfnamefont{I.}~\bibnamefont{Baldes}} \bibnamefont{and}
  \bibinfo{author}{\bibfnamefont{K.}~\bibnamefont{Petraki}},
  \bibinfo{journal}{JCAP} \textbf{\bibinfo{volume}{09}}, \bibinfo{pages}{028}
  (\bibinfo{year}{2017}), \eprint{1703.00478}.

\bibitem[{\citenamefont{Gresham
  et~al.}(2018{\natexlab{a}})\citenamefont{Gresham, Lou, and
  Zurek}}]{Gresham:2017cvl}
\bibinfo{author}{\bibfnamefont{M.~I.} \bibnamefont{Gresham}},
  \bibinfo{author}{\bibfnamefont{H.~K.} \bibnamefont{Lou}}, \bibnamefont{and}
  \bibinfo{author}{\bibfnamefont{K.~M.} \bibnamefont{Zurek}},
  \bibinfo{journal}{Phys. Rev. D} \textbf{\bibinfo{volume}{97}},
  \bibinfo{pages}{036003} (\bibinfo{year}{2018}{\natexlab{a}}),
  \eprint{1707.02316}.

\bibitem[{\citenamefont{HajiSadeghi et~al.}(2019)\citenamefont{HajiSadeghi,
  Smolenski, and Wudka}}]{HajiSadeghi:2017zrl}
\bibinfo{author}{\bibfnamefont{S.}~\bibnamefont{HajiSadeghi}},
  \bibinfo{author}{\bibfnamefont{S.}~\bibnamefont{Smolenski}},
  \bibnamefont{and} \bibinfo{author}{\bibfnamefont{J.}~\bibnamefont{Wudka}},
  \bibinfo{journal}{Phys. Rev. D} \textbf{\bibinfo{volume}{99}},
  \bibinfo{pages}{023514} (\bibinfo{year}{2019}), \eprint{1709.00436}.

\bibitem[{\citenamefont{Murase and Shoemaker}(2016)}]{Murase:2016nwx}
\bibinfo{author}{\bibfnamefont{K.}~\bibnamefont{Murase}} \bibnamefont{and}
  \bibinfo{author}{\bibfnamefont{I.~M.} \bibnamefont{Shoemaker}},
  \bibinfo{journal}{Phys. Rev. D} \textbf{\bibinfo{volume}{94}},
  \bibinfo{pages}{063512} (\bibinfo{year}{2016}), \eprint{1606.03087}.

\bibitem[{\citenamefont{Gresham
  et~al.}(2018{\natexlab{b}})\citenamefont{Gresham, Lou, and
  Zurek}}]{Gresham:2018anj}
\bibinfo{author}{\bibfnamefont{M.~I.} \bibnamefont{Gresham}},
  \bibinfo{author}{\bibfnamefont{H.~K.} \bibnamefont{Lou}}, \bibnamefont{and}
  \bibinfo{author}{\bibfnamefont{K.~M.} \bibnamefont{Zurek}},
  \bibinfo{journal}{Phys. Rev. D} \textbf{\bibinfo{volume}{98}},
  \bibinfo{pages}{096001} (\bibinfo{year}{2018}{\natexlab{b}}),
  \eprint{1805.04512}.

\bibitem[{\citenamefont{Lopes and Lopes}(2019)}]{Lopes:2019jca}
\bibinfo{author}{\bibfnamefont{J.}~\bibnamefont{Lopes}} \bibnamefont{and}
  \bibinfo{author}{\bibfnamefont{I.}~\bibnamefont{Lopes}},
  \bibinfo{journal}{Astrophys. J.} \textbf{\bibinfo{volume}{879}},
  \bibinfo{pages}{50} (\bibinfo{year}{2019}), \eprint{1907.05785}.

\bibitem[{\citenamefont{Ghosh et~al.}(2020)\citenamefont{Ghosh, Ghosh, and
  Mukhopadhyay}}]{Ghosh:2020lma}
\bibinfo{author}{\bibfnamefont{A.}~\bibnamefont{Ghosh}},
  \bibinfo{author}{\bibfnamefont{D.}~\bibnamefont{Ghosh}}, \bibnamefont{and}
  \bibinfo{author}{\bibfnamefont{S.}~\bibnamefont{Mukhopadhyay}},
  \bibinfo{journal}{JHEP} \textbf{\bibinfo{volume}{08}}, \bibinfo{pages}{149}
  (\bibinfo{year}{2020}), \eprint{2004.07705}.

\bibitem[{\citenamefont{Ghosh et~al.}(2021)\citenamefont{Ghosh, Ghosh, and
  Mukhopadhyay}}]{Ghosh:2021qbo}
\bibinfo{author}{\bibfnamefont{A.}~\bibnamefont{Ghosh}},
  \bibinfo{author}{\bibfnamefont{D.}~\bibnamefont{Ghosh}}, \bibnamefont{and}
  \bibinfo{author}{\bibfnamefont{S.}~\bibnamefont{Mukhopadhyay}},
  \bibinfo{journal}{Phys. Rev. D} \textbf{\bibinfo{volume}{104}},
  \bibinfo{pages}{123543} (\bibinfo{year}{2021}), \eprint{2103.14009}.

\bibitem[{\citenamefont{Ho et~al.}(2022)\citenamefont{Ho, Ko, and
  Lu}}]{Ho:2022erb}
\bibinfo{author}{\bibfnamefont{S.-Y.} \bibnamefont{Ho}},
  \bibinfo{author}{\bibfnamefont{P.}~\bibnamefont{Ko}}, \bibnamefont{and}
  \bibinfo{author}{\bibfnamefont{C.-T.} \bibnamefont{Lu}},
  \bibinfo{journal}{JHEP} \textbf{\bibinfo{volume}{03}}, \bibinfo{pages}{005}
  (\bibinfo{year}{2022}), \eprint{2201.06856}.

\bibitem[{\citenamefont{Steigerwald et~al.}(2022)\citenamefont{Steigerwald,
  Marra, and Profumo}}]{Steigerwald:2022pjo}
\bibinfo{author}{\bibfnamefont{H.}~\bibnamefont{Steigerwald}},
  \bibinfo{author}{\bibfnamefont{V.}~\bibnamefont{Marra}}, \bibnamefont{and}
  \bibinfo{author}{\bibfnamefont{S.}~\bibnamefont{Profumo}},
  \bibinfo{journal}{Phys. Rev. D} \textbf{\bibinfo{volume}{105}},
  \bibinfo{pages}{083507} (\bibinfo{year}{2022}), \eprint{2203.09054}.

\bibitem[{\citenamefont{Ho}(2022)}]{Ho:2022tbw}
\bibinfo{author}{\bibfnamefont{S.-Y.} \bibnamefont{Ho}},
  \bibinfo{journal}{JHEP} \textbf{\bibinfo{volume}{10}}, \bibinfo{pages}{182}
  (\bibinfo{year}{2022}), \eprint{2207.13373}.

\bibitem[{\citenamefont{Roy et~al.}(2024)\citenamefont{Roy, Dasgupta, and
  Guchait}}]{Roy:2024ear}
\bibinfo{author}{\bibfnamefont{A.}~\bibnamefont{Roy}},
  \bibinfo{author}{\bibfnamefont{B.}~\bibnamefont{Dasgupta}}, \bibnamefont{and}
  \bibinfo{author}{\bibfnamefont{M.}~\bibnamefont{Guchait}}
  (\bibinfo{year}{2024}), \eprint{2402.17265}.

\bibitem[{\citenamefont{Dutta~Banik}(2025)}]{DuttaBanik:2024vro}
\bibinfo{author}{\bibfnamefont{A.}~\bibnamefont{Dutta~Banik}},
  \bibinfo{journal}{Nucl. Phys. B} \textbf{\bibinfo{volume}{1013}},
  \bibinfo{pages}{116855} (\bibinfo{year}{2025}), \eprint{2408.01955}.

\bibitem[{\citenamefont{Su et~al.}(2024)\citenamefont{Su, Wu, and
  Yang}}]{Su:2024flx}
\bibinfo{author}{\bibfnamefont{L.}~\bibnamefont{Su}},
  \bibinfo{author}{\bibfnamefont{L.}~\bibnamefont{Wu}}, \bibnamefont{and}
  \bibinfo{author}{\bibfnamefont{M.}~\bibnamefont{Yang}},
  \bibinfo{journal}{Phys. Rev. D} \textbf{\bibinfo{volume}{110}},
  \bibinfo{pages}{055014} (\bibinfo{year}{2024}), \eprint{2408.03759}.

\bibitem[{\citenamefont{Wang}(2025)}]{Wang:2025ztb}
\bibinfo{author}{\bibfnamefont{J.-W.} \bibnamefont{Wang}}
  (\bibinfo{year}{2025}), \eprint{2503.22105}.

\bibitem[{\citenamefont{Baratella et~al.}(2014)\citenamefont{Baratella,
  Cirelli, Hektor, Pata, Piibeleht, and Strumia}}]{Baratella:2013fya}
\bibinfo{author}{\bibfnamefont{P.}~\bibnamefont{Baratella}},
  \bibinfo{author}{\bibfnamefont{M.}~\bibnamefont{Cirelli}},
  \bibinfo{author}{\bibfnamefont{A.}~\bibnamefont{Hektor}},
  \bibinfo{author}{\bibfnamefont{J.}~\bibnamefont{Pata}},
  \bibinfo{author}{\bibfnamefont{M.}~\bibnamefont{Piibeleht}},
  \bibnamefont{and} \bibinfo{author}{\bibfnamefont{A.}~\bibnamefont{Strumia}},
  \bibinfo{journal}{JCAP} \textbf{\bibinfo{volume}{03}}, \bibinfo{pages}{053}
  (\bibinfo{year}{2014}), \eprint{1312.6408}.

\end{thebibliography}

\end{document}